\documentclass[times,twocolumn,final]{elsarticle}
\usepackage{medima}
\usepackage{framed,multirow}

\usepackage{lineno,hyperref}
\modulolinenumbers[5]
\usepackage{amsmath}
\usepackage{amssymb}
\usepackage{verbatim}
\usepackage{multirow}
\usepackage{color,graphicx}
\usepackage{arydshln}
\usepackage{hyperref}
\usepackage{url}
\usepackage{footnote}
\usepackage{multirow}
\usepackage{bm}
\usepackage{ulem}
\usepackage{cite}
\usepackage{subfigure}

\providecommand{\Zxhrefeq}[1]{Equation (\ref{#1})}
\providecommand{\Zxhreftb}[1]{Table~\ref{#1}}

\providecommand{\zxhreffig}[1]{Fig.~\ref{#1}}
\providecommand{\Zxhreffig}[1]{Fig.~\ref{#1}}

\modulolinenumbers[5]

\journal{Medical Image Analysis}

\begin{document}

\begin{frontmatter}
\title{AtrialJSQnet: A New Framework for Joint Segmentation and Quantification of Left Atrium and Scars Incorporating Spatial and Shape Information} %Incorporating Spatial and Shape Information 

\author[label1,label2,label3]{Lei Li} %lilei.sky@sjtu.edu.cn
\author[label3,label4]{Veronika A. Zimmer} %veronika.zimmer@kcl.ac.uk
\author[label3,label4,label5]{Julia A. Schnabel} %julia.schnabel@kcl.ac.uk
\author[label1]{Xiahai Zhuang*} %zxh@fudan.edu.cn
\ead[url]{zxh@fudan.edu.cn}%; homepage: http://www.sdspeople.fudan.edu.cn/zhuangxiahai}

\address[label1]{School of Data Science, Fudan University, Shanghai, China}
\address[label2]{School of Biomedical Engineering, Shanghai Jiao Tong University, Shanghai, China}
\address[label3]{School of Biomedical Engineering and Imaging Sciences, King’s College London, London, UK}
\address[label4]{Technical University Munich, Munich, Germany}
\address[label5]{Helmholtz Center Munich, Germany}

\received{11 Aug 2020}
% \revised{8 Oct 2021}
\accepted{8 Nov 2021}
% \availableonline{1 June 2021}
% \communicated{S. Sarkar}

\begin{abstract}
Left atrial (LA) and atrial scar segmentation from late gadolinium enhanced magnetic resonance imaging (LGE MRI) is an important task in clinical practice. %, to guide ablation therapy and predict treatment results for atrial fibrillation (AF) patients. 
The automatic segmentation is however still challenging due to the poor image quality, the various LA shapes, the thin wall, and the surrounding enhanced regions.
Previous methods normally solved the two tasks independently and ignored the intrinsic spatial relationship between LA and scars.
In this work, we develop a new framework, namely AtrialJSQnet, where LA segmentation, scar projection onto the LA surface, and scar quantification are performed simultaneously in an end-to-end style.
We propose a mechanism of shape attention (SA) via an implicit surface projection to utilize the inherent correlation between LA cavity and scars.
In specific, the SA scheme is embedded into a multi-task architecture to perform joint LA segmentation and scar quantification.
Besides, a spatial encoding (SE) loss is introduced to incorporate continuous spatial information of the target in order to reduce noisy patches in the predicted segmentation.
%Moreover, the proposed method can alleviate the severe class-imbalance problem when detecting small and discrete scars. 
We evaluated the proposed framework on 60 post-ablation LGE MRIs from the \textit{MICCAI2018 Atrial Segmentation Challenge}.
Moreover, we explored the domain generalization ability of the proposed AtrialJSQnet on 40 pre-ablation LGE MRIs from this challenge and 30 post-ablation multi-center LGE MRIs from another challenge (\textit{ISBI2012 Left Atrium Fibrosis and Scar Segmentation Challenge}).
% For LA segmentation, the proposed method reduced the mean Hausdorff distance from 36.4 mm to 20.0 mm, compared to the 3D U-Net using the binary cross-entropy loss.
% For scar quantification, the method was compared with the results or algorithms reported in the literature and demonstrated a comparative performance.
Extensive experiments on public datasets demonstrated the effect of the proposed AtrialJSQnet, which achieved competitive performance over the state-of-the-art. 
The relatedness between LA segmentation and scar quantification was explicitly explored and has shown significant performance improvements for both tasks. 
The code has been released via https://github.com/Marie0909/AtrialJSQnet.

\end{abstract}


\end{frontmatter}

%\linenumbers

\section{Introduction}
Atrial fibrillation (AF) is the most common cardiac arrhythmia in the clinic, especially in the aged population \citep{journal/cir/chugh2014}.
Late gadolinium enhanced magnetic resonance imaging (LGE MRI) has been widely used to visualize the extent and distribution of scars.
The segmentation and quantification of left atrial (LA) and scars from LGE MRI provide reliable information for patient selection, clinical diagnosis, and treatment stratification \citep{journal/EP/njoku2018}.
% Radiofrequency ablation is a promising procedure for treating AF, where patient selection and outcome prediction of such therapy can be improved through left atrial (LA) scar localization and quantification. 
% Atrial scars are located on the LA wall, thus it normally requires LA/ LA wall segmentation to exclude confounding enhanced tissues from other substructures of the heart, as \zxhreffig{fig:method:workflow} shows.
% challenge of auto seg
Manual delineations of LA and scars are time-consuming and prone to be subjective, so automatic segmentation is highly desired.
However, the development of automatic techniques remains challenging, mainly due to poor image quality, various LA shapes, thin LA walls and enhanced noise from surrounding tissues.

\begin{figure}[t]\center
 \includegraphics[width=0.45\textwidth]{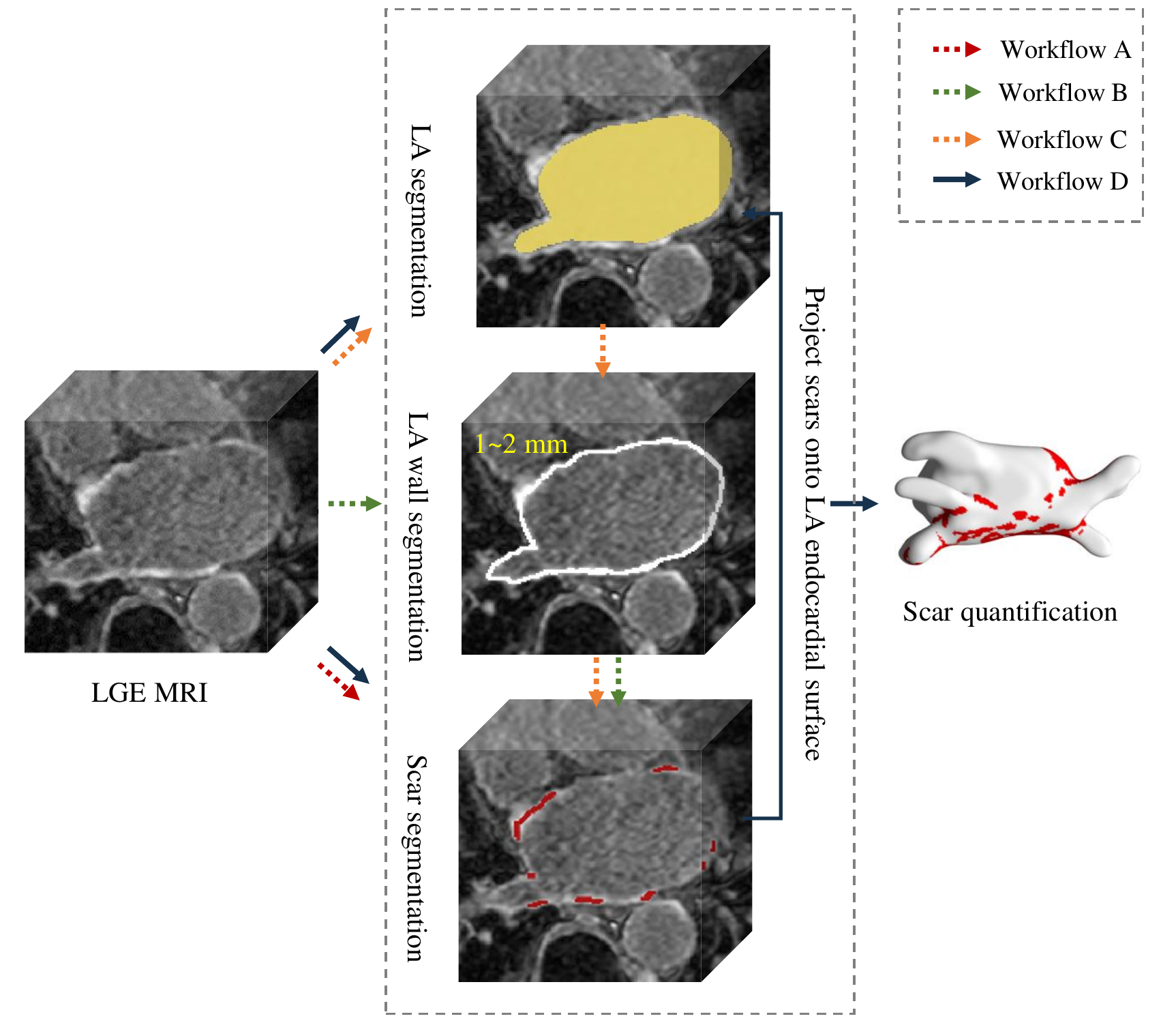}\\[-2ex]
   \caption{Four kinds of workflows for the volume-based scar segmentation or surface-based scar quantification. Here, the target LGE MRI is cropped for better visualization.}
\label{fig:method:workflow}\end{figure}

Limited studies have been reported in the literature to develop joint LA segmentation and scar quantification algorithms.
In fact, for scar segmentation/ quantification, most of the current works require an accurate initialization of LA or LA wall, as scars are located on the LA wall \citep{conf/MI/perry2012,journal/jcmr/Karim2013,journal/TEHM/karim2014}.
% adopted threshold-based methods that relied on manual LA wall segmentation \citep{journal/jcmr/Karim2013}.
% Other conventional algorithms, such as Gaussian mixture model (GMM) \citep{journal/TEHM/karim2014} and k-means \citep{conf/MI/perry2012} 
\zxhreffig{fig:method:workflow} summarizes four workflows for scar segmentation/ quantification.
First, one can directly segment scars from LGE MRI, i.e., workflow A, but it could be challenging due to the small volume of scars and enhanced adjacent regions.
Second, one can firstly segment the LA wall, and subsequently classify each segmented voxel as either normal wall or scars, i.e., workflow B.
However, automatic LA wall segmentation remains a challenging problem due to its inherent thin thickness  \citep{journal/MedAI/karim2018}. %($1\sim2$ mm)  (under 2 mm)
Third, the most common procedure is to first segment the LA, and then generate the wall segmentation by defining a fixed distance from the LA endocardium \citep{journal/jcmr/Karim2013}; %, such as using Euclidean distance or morphological dilation; or using active contour to estimate the LA epicardium;
Based on the approximated wall segmentation, scars can be extracted inside it, i.e., workflow C.

Clinical studies mainly focus on the location and extent of scars, suggesting that wall thickness can be ignored \citep{journal/tmi/Ravanelli2014,journal/MedIA/li2020,conf/STACOM/li2018}.
Therefore, one can skip the step of wall segmentation, and directly project the segmented scar onto the LA surface for scar quantification, i.e., workflow D.
For example, a maximum intensity projection technique \citep{conf/STACOM/qiao2018} and a 2D skeleton algorithm \citep{journal/tmi/Ravanelli2014} were employed to project scars onto the LA surface.
Recently, \citet{journal/MedIA/li2020} proposed a graph-cuts framework for scar quantification on the LA surface mesh, where weights of a graph were learned via a multi-scale convolutional neural network (CNN).
They achieved LA segmentation and scar quantification independently, instead of considering the two tasks in a unified framework. %, such as the work of \citet{conf/MICCAI/chen2018}.
To utilize the spatial relationship between LA and scars, they extracted multi-scale patches (MSP) with random offsets along the perpendicular direction of the LA endocardial surface, which could eliminate the effect of inaccurate LA segmentation.
However, the MSP strategy resulted in an expensive time and space complexity, so they could not achieve end-to-end training with optimization on the whole graph.
%Therefore, it did not fully utilize the learning capability of neural networks.

%our proposals
In this work, we present a new framework based on deep neural networks, which can jointly perform the segmentation and quantification of blood cavity and scars of LA from LGE MRI, and therefore is referred to as AtrialJSQnet.
%end-to-end joint segmentation and quantification network for LA and scars via multi-task learning, namely AtrialJSQnet.
To explicitly utilize the spatial relationship between LA and scars, we adopt the LA surface as an attention mask on the predicted scar probability map for shape attention (SA).
In this way, scars are also projected onto the LA surface and therefore circumvent the challenging task of wall segmentation.
Besides, we introduce a novel spatially encoded (SE) loss, which incorporates spatial information in the pipeline without any modifications of the networks.
The SE loss relies on the distance transform map (DTM) \citep{journal/PAMI/1995} (equivalent to signed distance functions), which is generated from the binary label and can be regarded as a continuous representation of the ground truth. 
%to eliminate outliers for LA segmentation, with additional benefits for scar quantification.  
% Both shape attention module and spatial encoding scheme are embedded into a multi-task network.
Therefore, AtrialJSQnet incorporates both spatial and shape information, and achieves the simultaneous LA segmentation, scar projection onto the LA surface and scar quantification.
This paper significantly extends a preliminary conference version of the work presented in \citep{conf/MICCAI/li2020}.
Specifically, a range of additional experiments are carried out to verify the performance of the proposed SE module for LA segmentation.
A model generalization study is included by testing the trained model (training data is post-ablation LGE MRI) on the pre-ablation LGE MRI and multi-center LGE MRI dataset.
In addition, a more extensive analysis and discussion are added on both LA segmentation and scar quantification results.
The remainder of the paper is organized as follows: relevant studies are introduced in Section \ref{literature}. 
The detailed framework of the proposed algorithm is presented in Section \ref{method}. 
Section \ref{exp} presents the experiments and results. 
Discussion and conclusion are given in Section \ref{discussion}.

\section{Related work} \label{literature}
Besides scar quantification, this work is also related to the following three topics: (1) LA and LA wall segmentation, (2) shape regularization in deep learning (DL)-based segmentation, and (3) multi-task learning.

\begin{figure*}[t]\center
 \includegraphics[width=0.9\textwidth]{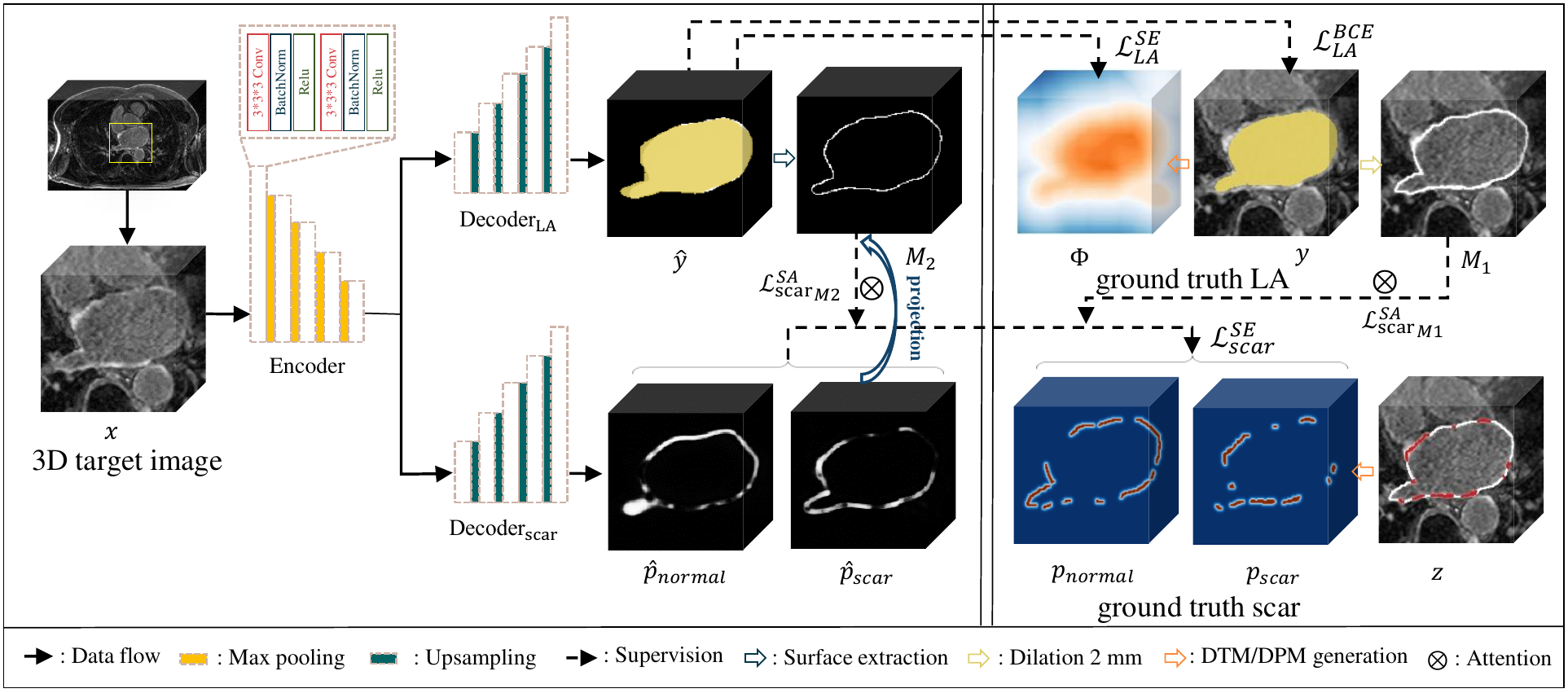}\\[-2ex]
   \caption{The proposed AtrialJSQnet framework for joint LA segmentation and scar quantification. Note that the skip connections between the encoder and two decoders are omitted here for simplification. The reader is referred to Section \ref{method} for the explanation of the symbols in this diagram.}
\label{fig:method:network}
\end{figure*}

\subsection{LA and LA wall segmentation} 
For LA segmentation, \citet{journal/tmi/tobon2015} reported the results of nine algorithms in the Left Atrial Segmentation Challenge 2013 (LASC 2013), which provided 30 CT and 30 MRI datasets for LA segmentation.
The evaluated algorithms in this challenge were mostly model or atlas-based methods, such as region growing, statistical shape models (SSM), and multi-atlas segmentation (MAS).
However, it could be difficult to obtain a reasonable result when applying these methods to LGE MRI directly, as LGE MRI has relatively poor quality in general.
A common way to solve this problem is to combine LGE MRI with additional images, such as the balanced steady-state free precession (bSSFP) MRI, to incorporate shape prior \citep{journal/tmi/Ravanelli2014,journal/jmri/Tao2016,journal/MP/yang2018,journal/MedIA/li2020}.
Recently, a new LA segmentation challenge providing 154 LGE MRIs was organized, where many DL-based methods were employed to directly segment LA from LGE MRI \citep{journal/xiong2020}.
For example, \citet{conf/STACOM/chen2018} presented a two-task network for LA segmentation and patient classification.% to incorporate the prior information of the patient category.
\citet{conf/STACOM/yang2018} designed a deep network with transfer learning and employed a deep supervision strategy for LA segmentation.
% \citet{conf/MICCAI/yu2019} designed an uncertainty-aware semi-supervised framework for LA segmentation and validated their method on this dataset.
In this challenge, the results of DL-based methods were significantly better than that of traditional atlas-based methods ($p<0.05$) \citep{journal/xiong2020}. 
 
For LA wall segmentation, \citet{journal/mia/Veni2017} proposed an algorithm, namely ShapeCut, combining a shape-based system and graph-cuts approach for a dual surface estimation.
\citet{conf/miccai/wu2018} adopted a multivariate mixture model (MvMM) and the maximum likelihood estimator for the LA wall and scar segmentation by combining LGE MRI and bSSFP MRI.
\citet{conf/ip/Ji2018} utilized the advanced two-layer level set for dual surface segmentation of LA and LV wall based on a manual initialization.
%Their method was 2D-based and required a manual initialization of the endocardial boundaries.
\citet{journal/MedAI/karim2018} presented the submitted results from the Segmentation of Left Atrial Wall for Thickness (SLAWT 2016) challenge, which provided 10 CT and 10 MRI public datasets.
Due to the difficulty of the task, the participants only submitted results on CT, and the challenge organizers had to evaluate their proposed three algorithms on MRI for a complete benchmark.
% There was no clear winner in CT, but the level-set performed best compared to other techniques in MRI.
Therefore, in this work we circumvent the challenging task of LA wall segmentation, and project scars onto the corresponding LA surface for scar quantification.

\subsection{Shape regularization in DL-based segmentation}
%shape constrain
DL-based methods have shown significant advantages in medical image segmentation tasks.
However, most neural networks are trained with a loss only considering the label mask in a discrete space, which could fail to learn high-level topological shape information.
Due to the lack of spatial information, predictions often include large outliers or unrealistic shapes \citep{journal/MedAI/zhuang2019,journal/PAMI/zhuang2019}. 
%commonly tend to be with large outliers or unrealistic shapes.
To tackle this issue, some strategies have been utilized, such as graph-cuts/ conditional random field (CRF) regularization \citep{journal/MedIA/li2020,journal/MedAI/kamnitsas2017}, deep level set \citep{conf/tang2017}, deformable-model based shape refinement \citep{journal/MedAI/avendi2016,journal/TMI/duan2019,journal/TMI/lee2019,conf/MICCAI/zeng2019}, shape reconstruction \citep{journal/TMI/oktay2017,conf/MICCAI/yue2019} and SSM \citep{journal/TMI/mansoor2016,journal/JCARS/karimi2018}. %,journal/arXiv/clough2019
However, the above studies were not trained in an end-to-end fashion or required heavy network modification.

Recently, distance map regularized networks were proposed to predict segmentations and corresponding distance maps in a two-task style.
The utilization of distance maps can force the network to learn more concrete spatial structural information, than just classifying a pixel/voxel lying inside or outside of a binary target mask.
For example, \citet{journal/MP/dangi2019} designed a two-task network for semantic segmentation and pixel-wise distance map regression.
\citet{journal/CVIU/audebert2019} employed a network to first predict the DTM, and then used an additional convolutional layer to fuse the inferred DTM and the last layer features, to perform the final segmentation. 
Instead of adding additional layers, \citet{conf/xue2019} directly predicted the segmentation by employing an approximated Heaviside function on the inferred DTM.
\citet{journal/TMI/karimi2019} proposed three strategies to estimate the Hausdorff distance (HD) from the segmentation probability map generated by CNN, including distance transform, morphological erosion, and circular/ spherical convolution kernels. 
They aimed to directly reduce HD by using the newly designed HD-based loss functions for shape regularization.
However, their method is computationally expensive, due to the computing of the distance transforms in the back-propagation.
By contrast, in this work we introduce a DTM-based spatial encoding loss function, which can be optimized efficiently in an end-to-end fashion without any modifications to the networks.

\begin{figure*}[t]\center
    \subfigure[] {\includegraphics[width=0.17\textwidth]{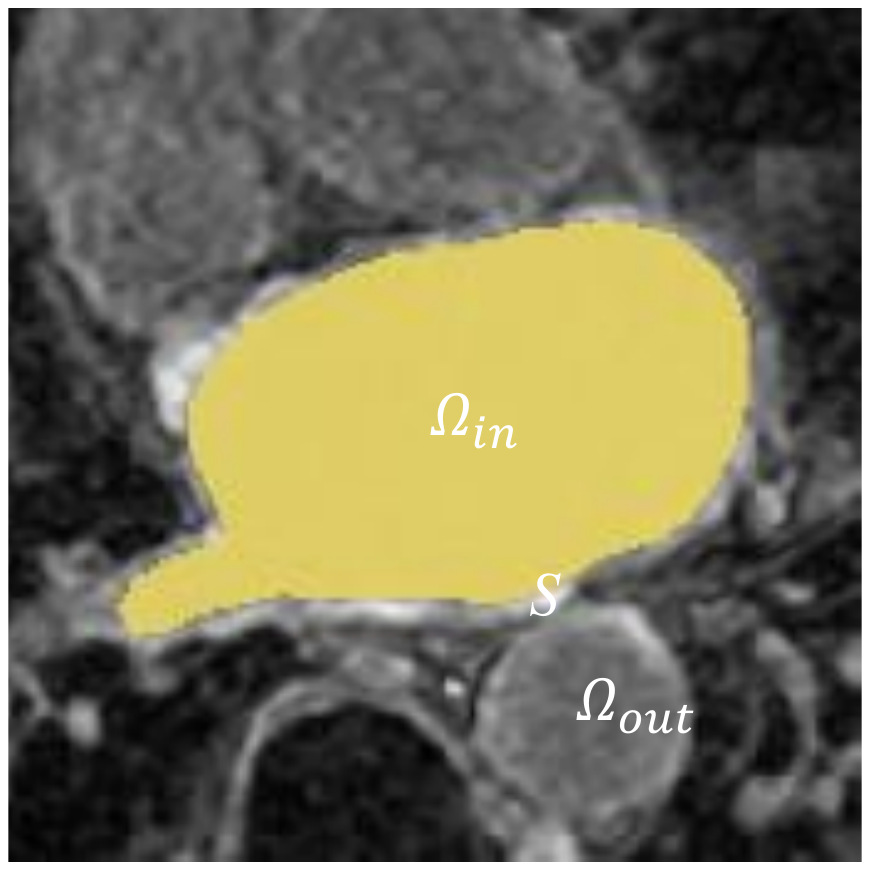}}
    \subfigure[] {\includegraphics[width=0.2\textwidth]{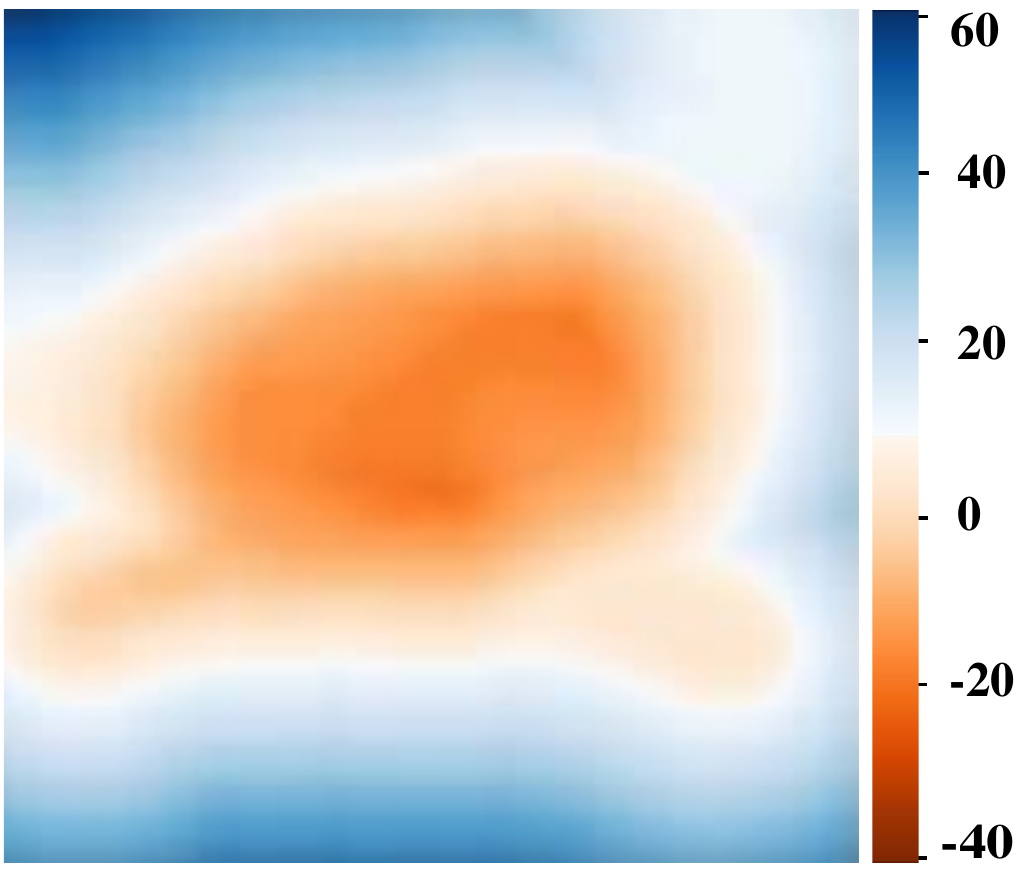}}
    \subfigure[] {\includegraphics[width=0.17\textwidth]{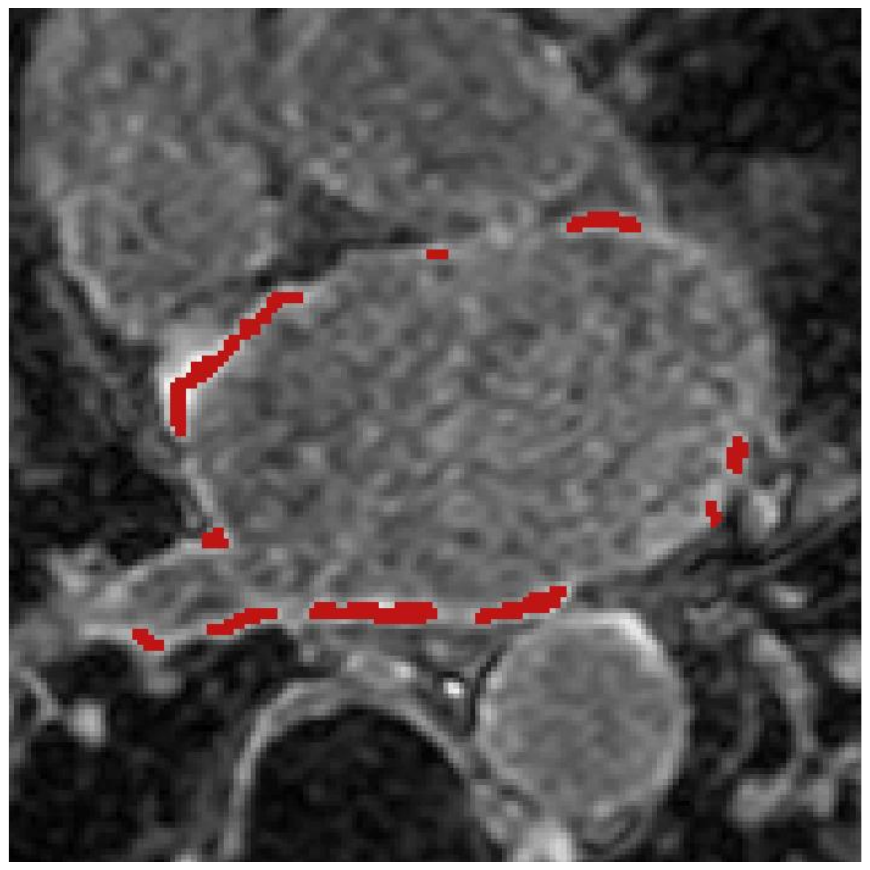}}
    \subfigure[] {\includegraphics[width=0.17\textwidth]{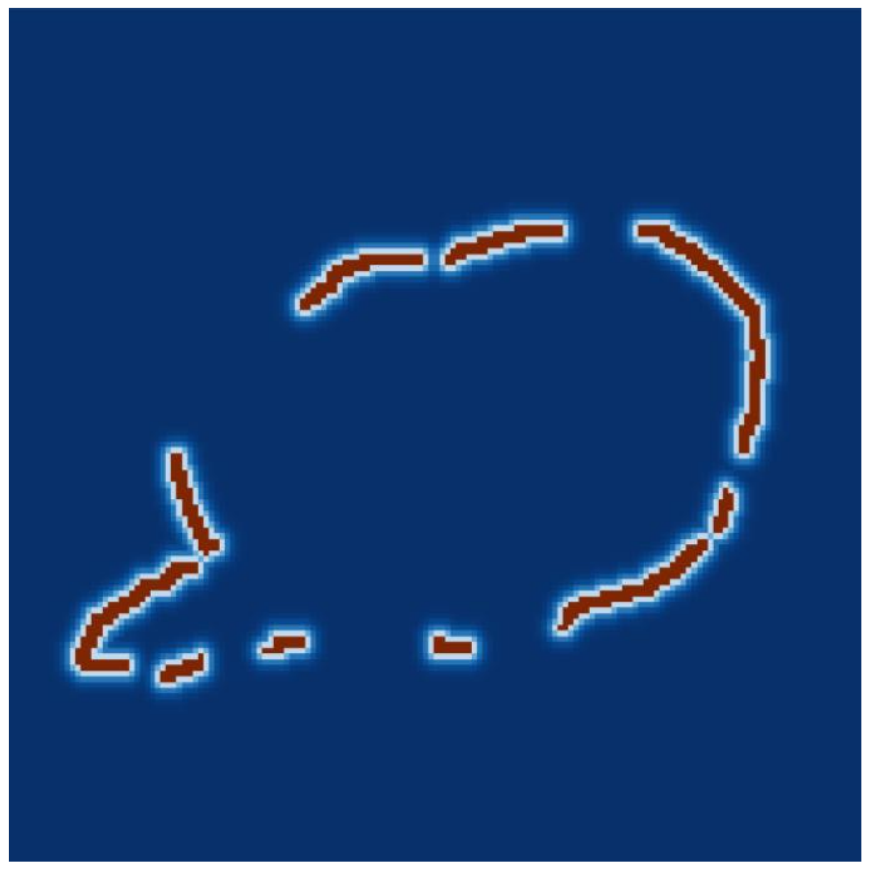}}
    \subfigure[] {\includegraphics[width=0.2\textwidth]{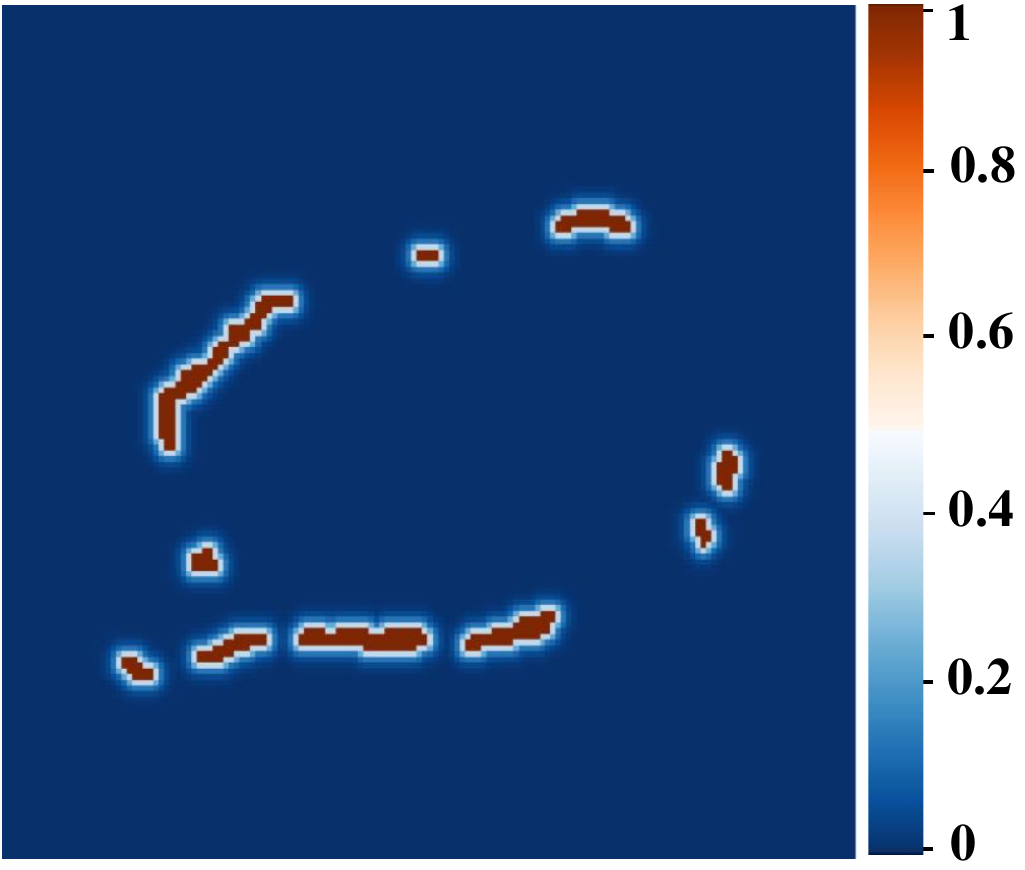}}
   \caption{Presentations of ground truth for LA segmentation and scar quantification:
     (a) binary LA label;
     (b) signed DTM of LA with a color-bar referring to the signed distance value); 
     (c) binary label of scars (scarring regions on the wall);
     (d) DPM of normal wall (healthy regions on the wall);
     (e) DPM of scars with a color-bar indicating the probability value of normal wall/ scars.
      }
\label{fig:method:SDM}\end{figure*}

\subsection{Multi-task learning}
Joint/ simultaneous segmentation, quantification or classification is generally implemented via multi-task learning (MTL), which has been shown to outperform methods considering related tasks separately.
%Normally, it can introduce other priors and sometimes achieve an attention scheme. 
\citet{journal/MedAI/zhang2020} proposed a context-guided multi-task CNN for both craniomaxillofacial bone segmentation and landmark digitization. 
To capture the spatial context information of the image, they employed displacement maps for modeling the displacement information between voxels and landmarks.
\citet{conf/MICCAI/chen2019} employed a multi-task attention-based network and achieved semi-supervised learning by simultaneously optimizing a supervised segmentation and an unsupervised reconstruction task.
\citet{conf/MICCAI/xue2017} proposed an MTL network to model the intra- and inter-task relatedness for left ventricle quantification.
\citet{journal/tmi/liu2019} developed a multi-task DL network to model the relationship between lung nodule classification and attribute score regression.
\citet{conf/MICCAI/chen2018} and \citet{journal/FGCS/yang2020} employed a multiview two-task (MvTT) recursive attention model to segment LA and scars simultaneously.
Although MTL has proved to be effective, it is crucial to find and effectively incorporate the relationship between multiple tasks.
In this work, we propose to explicitly learn the spatial relationship of LA and scars via a shape attention scheme combined with a surface projection.

\section{Method} \label{method}
\zxhreffig{fig:method:network} provides an overview of the proposed framework, i.e., AtrialJSQnet (abbreviated as AJSQnet).
AtrialJSQnet is a modified U-Net consisting of two decoders for LA segmentation and scar quantification, respectively.
For LA segmentation, an SE loss based on the DTM is introduced as a shape regularization term to avoid noisy segmentation (see Section \ref{method:LA}).
For scar segmentation, an SE loss is also employed with a surface projection (see Section \ref{method:scar}).
To utilize the spatial relationship between LA and scar, a specific SA scheme is embedded in the AtrialJSQnet (see Section \ref{method:multi_task}).
The SA scheme also ensures that the surface projection is realized in an end-to-end style.

\subsection{Spatially encoded constraint for LA segmentation}\label{method:LA}
To exclude confounding enhanced tissues from other substructures of the heart, LA segmentation is required for scar quantification.
DL-based methods, especially U-Net, have achieved promising performance for cardiac image segmentation \citep{journal/MedAI/zhuang2019}.
However, one common shortage of current DL-based methods is the lack of spatial awareness which often leads to noisy segmentation, especially for highly variable structures such as the LA.
This is mainly because in these methods all voxels are independently classified without considering their spatial relation to the shape of the target structure, compared to conventional model-based or atlas-based methods.

We propose a distance-based SE loss as a regularization term for LA segmentation, which assigns different penalties to false classifications in different positions.
For example, the outliers far from the target will be heavily penalized.
In this way, the network is supposed to learn more information about the underlying spatial structure of the target compared to traditional DL-based classification.
Given a target label, the signed DTM for each voxel $x_i$ can be defined as,
\begin{equation}
\phi(x_i)=\begin{cases} -d^\beta & x_i \in \Omega_{in} \\0 & x_i \in S\\ d^\beta & x_i \in \Omega_{out} \end{cases}
\label{eq:SDM}
\end{equation}
where $\Omega_{in}$ and $\Omega_{out}$ respectively indicate the region inside and outside the target label, $S$ denotes the surface boundary, $d$ represents the distance from voxel $x_i$ to the nearest point on $S$, and $\beta$ is a hyperparameter.
\zxhreffig{fig:method:SDM} (a) and (b) show the target LA label and its signed DTM, which can be regarded as a discrete and continuous representation of ground truth, respectively.
Note that we clip the distance with a threshold equals to 50 to avoid large-range spatial computation.
% The binary cross-entropy (BCE) loss and the additional SE loss for LA segmentation can be defined as,
The loss for LA segmentation can be defined as,
\begin{equation}
  \mathcal L_{LA} = \mathcal L_{LA}^{BCE} + \lambda_{LA}\mathcal L_{LA}^{SE},
\end{equation}
where
\begin{equation}
  \mathcal L_{LA}^{BCE} = \sum_{i=1}^N y_i \cdot \log(\hat{y}(x_i; \theta)) + (1-y_i) \cdot \log(1-\hat{y}(x_i; \theta)),
\end{equation}
and
\begin{equation}
  \mathcal L_{LA}^{SE} = \sum_{i=1}^N (\hat{y}(x_i; \theta)-T_{LA}) \cdot \phi(x_i),
\end{equation}
where $\hat{y}$ and $y$ ($y\in\{0,1\}$) are the prediction of LA and its ground truth, respectively, $N$ is the number of voxels, $\lambda_{LA}$ is a balancing parameter, $T_{LA}$ is the threshold for LA segmentation, and $\cdot$ denotes the dot product.
Here, $T_{LA}$ is set to 0.5 to distinguish LA and background regions, i.e., $\Omega_{in}$ and $\Omega_{out}$.
For erroneous predictions including both false positives and false negatives, the value of $L_{LA}^{SE}$ will be positive as a penalty.
Therefore, the spatial information can be encoded by assigning different weights, i.e., $\phi(x_i)$, to each voxel according to its distance to the boundary of the target.

\zxhreffig{fig:method:LAerror} illustrates the conception. 
The penalty of the erroneous LA segmentation on point $A'$ should be considerably larger than that of point $B'$ and $C'$, because the distance of $AA'$ is much larger than $BB'$ and $CC'$.

% \subsection{Spatially encoded constraint with an implicit projection for scar quantification}  \label{method:scar}
\subsection{Implicit projection and spatially encoded constraint for scar quantification}  \label{method:scar}
To ignore the wall thickness which varies across different positions and different patients \citep{journal/MedAI/karim2018}, we propose to project the extracted scars onto the LA surface.
Here, the LA surface is represented by the boundary voxels of the LA segmentation, and can be extracted according to the gradient of the predicted LA probability maps.
Therefore, the volume-based scar segmentation problem is converted into the task of surface-based scar quantification through the implicit surface projection.
However, the voxel-wise classification in the surface-based quantification task only includes very limited information, i.e., the intensity value of one voxel. 
Besides, there usually exists misalignment between the extracted endocardial surface and ground truth due to errors or biases from the automatic LA segmentation.
In contrast to extracting multi-scale patches along the LA surface \citep{journal/MedIA/li2020}, we employ the SE loss to learn the spatial features near the LA surface.
Thus, it can be beneficial to improving the robustness of the framework against the LA segmentation errors.

We encode the spatial information by adopting the probability maps of normal wall and scarring regions as the ground truth, instead of the binary scar label $z$.
The probability map is generated from the DTM of normal wall or scars, and therefore is referred as distance probability map (DPM).
In contrast to traditional DL-based segmentation algorithms which optimize a discrete loss function, the DPM considers the continuous spatial information of scars and normal walls.
Specifically, we separately obtain the DTM of the scar and normal wall from a manual scar label, and convert them into probability maps $p(x_i)= [p_{normal}(x_i), p_{scar}(x_i)]$ with $p=exp^{-|\phi(x)|}$.
Note that as we do not consider the probability of the background (voxels that belong to neither normal wall nor scars), the probability maps generated here are not normalized.
Then, the SE loss for scar quantification can be defined as,
\begin{equation}
  \mathcal L_{scar}^{SE} = \sum_{i=1}^N \|\hat{p}(x_i; \theta) - p(x_i)\|^2_2,
\label{eq:SE_scar}
\end{equation}
where $\hat{p}$ ($\hat{p} = [\hat{p}_{normal}, \hat{p}_{scar}]$) is the predicted distance probability map of both normal wall and scarring region.
One can compare these two probabilities to decide whether the voxel in question belongs to a scar region or the normal wall, instead of employing a fixed threshold.
This is because the probability value can be unstable due to inaccurate LA segmentation. 
For example, the value will be very small when the voxel is far from the wall, as \zxhreffig{fig:method:SDM} (c-e) shows.
In contrast, the comparison of the two probabilities is relatively fixed along the perpendicular direction of the surface, which could mitigate the effect of inaccurate LA segmentation.

\begin{figure}[t]\center
 \includegraphics[width=0.48\textwidth]{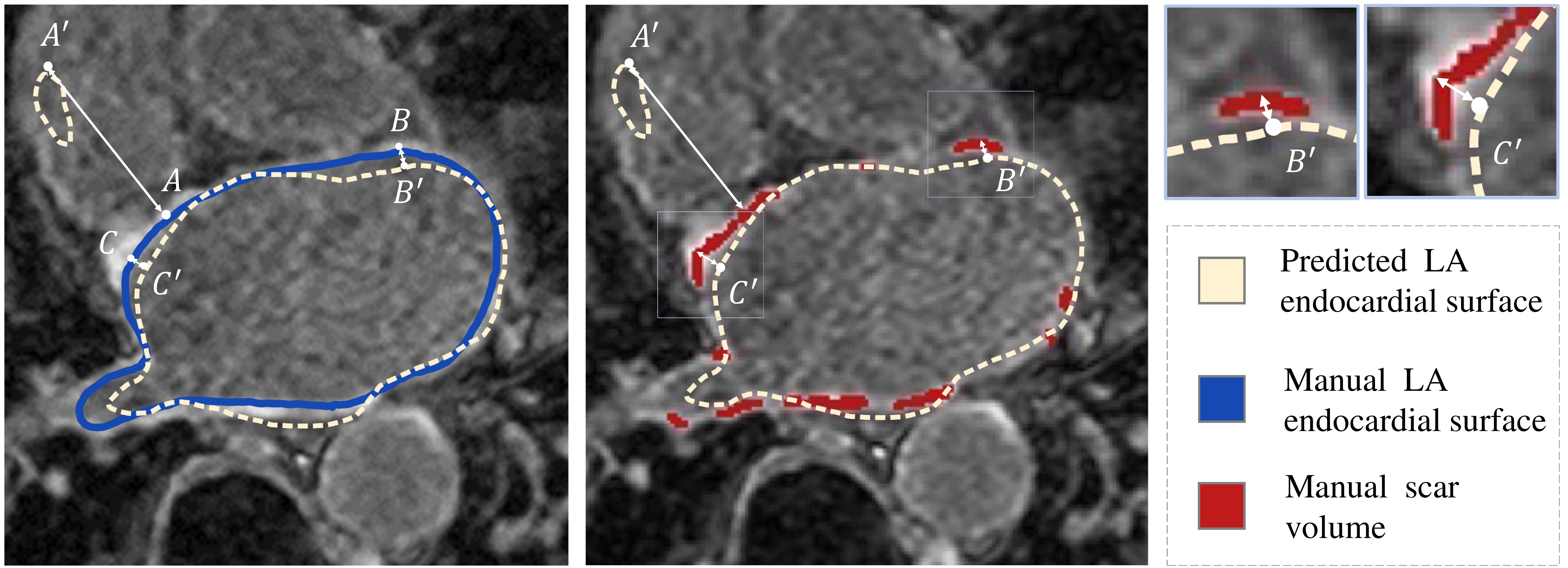}\\[-2ex]
   \caption{An example to illustrate the misalignment between predicted and ground truth LA endocardium  as well as its effect on scar quantification.
   Here, $A^{\prime}$, $B^{\prime}$ and $C^{\prime}$ are the points on the predicted LA endocardium, while $A$, $B$ and $C$ are their corresponding nearest points on the ground truth of LA endocardial surface.
   }
\label{fig:method:LAerror}\end{figure}

As \zxhreffig{fig:method:LAerror} shows, points $B^{\prime}$ and $C^{\prime}$ still could be classified into scars on the predicted LA surface, while not being inside the binary scar label.
Therefore, although there exists a small misalignment between the manual and predicted LA, one can still distinguish the scars.
However, large outliers in the LA segmentation, such as point $A^{\prime}$, will result in a wrong projection and adversely affect the accuracy of scar quantification.
%This is one of the reasons why we add spatial constrain on LA segmentation in Section~\ref{method:LA}.

\subsection{End-to-end trainable shape attention via multi-task learning} \label{method:multi_task}
To achieve simultaneous LA and atrial scar segmentation, we design a multi-task network including two decoders.
As \zxhreffig{fig:method:network} shows, the Decoder$_\text{LA}$ aims to predict the LA segmentation, and the Decoder$_\text{scar}$ infers the probability maps for scar quantification.
To explicitly learn the relationship between the two tasks, we extract the LA boundary from the predicted LA as an attention mask to achieve the implicit surface projection.
An SA loss is therefore introduced to enforce the attention of Decoder$_\text{scar}$ on the LA wall,
\begin{equation}
  \mathcal L^{SA}_{scar} = \sum_{i=1}^N (M \cdot (\nabla \hat{p}(x_i; \theta) - \nabla p(x_i)))^2,
\end{equation}
where $\nabla \hat{p} = \hat{p}_{normal} - \hat{p}_{scar}$, 
$\nabla p = p_{normal}-p_{scar}$, and $M$ is the attention mask, which can be generated from the gold standard segmentation of LA ($M_{1}$) as well as the predicted LA ($M_{2}$).
Here, we constrain the difference between $p_{normal}$ and $p_{scar}$ on the mask, to emphasize the comparison of the two probabilities.
Mask $M_{1}$ forces the predicted scar to locate inside the ground truth wall, while mask $M_{2}$ aims to connect the two tasks for spatial relationship learning and achieves an end-to-end projection.
Also, the attention scheme can alleviate the class-imbalance problem caused by the intrinsic characteristics of scars (scars are typically small regions and discrete).
The overall optimization problem for the proposed method is then formulated as follows,
\begin{equation}
   \mathcal L = \mathcal L_{LA} + \lambda_{scar}\mathcal L_{scar}^{SE}
        + \lambda_{M_{1}}\mathcal L^{SA}_{scarM_1} + \lambda_{M_{2}}\mathcal L^{SA}_{scarM_2},
\end{equation}
where $\lambda_{scar}$, $\lambda_{M_{1}}$ and $\lambda_{M_{2}}$ are balancing parameters.

\section{Experiments and results} \label{exp}
\subsection{Data acquisition and experiment setup}
%dataset
The public dataset used in this study is from \textit{MICCAI2018 Atrial Segmentation Challenge} \citep{link/LAseg2018,journal/TMI/xiong2018}.
The 100 LGE MRI training data from this challenge, consists of 60 post-ablation and 40 pre-ablation images with manual segmentation of LA.
In this work, we mainly focused on the 60 post-ablation data and manually segmented the LA scars for experiments.
All the LGE MRIs were reconstructed to $1\!\times\!1\!\times\!1$ mm, were cropped into a unified size of $208\!\times\!208\!\times\!80$ centering at the image and were normalized using Z-score.
We randomly split the data into training (40 subjects) and testing (20 subjects) subsets in experiments.

To evaluate the model generalization ability, we also included the 40 pre-ablation LGE MRIs from \textit{MICCAI2018 Atrial Segmentation Challenge} and additional 30 post-ablation LGE MRIs from \textit{ISBI2012 LA Fibrosis and Scar Segmentation Challenge} \citep{link/LAScarSeg2012}.
The dataset was collected from three different centers, i.e., University of Utah (Utah), Beth Israel Deaconess Medical Center (BIDMC), and King’s College London (KCL).
As presented in \Zxhreftb{tb:AtrialGeneral:data info}, the acquisition parameters differ between the centers.
The generalization study is reported in Sec \ref{result_DG}.

\begin{table*} [t] \center
    \caption{
    Image acquisition parameters of the two datasets, i.e., \textit{MICCAI2018 Atrial Segmentation Challenge} dataset and \textit{ISBI2012 Left Atrium Fibrosis and Scar Segmentation Challenge} dataset. 
    $I^{\text{pre}}$ and $I^{\text{post}}$ refer to pre- and post-ablation LGE MRI, respectively, and $^\P$ denotes the data from ISBI2012 challenge.
    Note that the pre-ablation LGE MRIs of ISBI2012 were not included in this study.
     }
\label{tb:AtrialGeneral:data info}
{\small
%\begin{tabular}{  l| l | l l l *{5}{@{\ \,} l }}
\begin{tabular}{p{1.8cm}|p{3.3cm}|p{3.2cm}|p{3.2cm}|p{3.2cm}} %No. subjects
\hline
\multirow{2}*{Parameters} & MICCAI2018 dataset & \multicolumn{3}{c}{ISBI2012 dataset}\\
\cline{2-5}
~    &Utah & Utah & BIDMC & KCL \\
\hline
No. subject &40 $I^{\text{pre}}$ + 60 $I^{\text{post}}$ & 10 $I^{\text{pre}\P}$ + 10 $I^{\text{post}\P}$  & 10 $I^{\text{pre}\P}$ + 10 $I^{\text{post}\P}$  & 10 $I^{\text{pre}\P}$ + 10 $I^{\text{post}\P}$ \\
\hdashline
Scanner & 1.5T Siemens Avanto; 3T Siemens Vario & 1.5T Siemens Avanto; 3T Siemens Vario & 1.5T Philips Achieva & 1.5T Philips Achieva  \\
\hdashline
Resolution &1.25 $\!\times\!$ 1.25 $\!\times\!$ 2.5 mm &1.25 $\!\times\!$ 1.25 $\!\times\!$ 2.5 mm & 1.4 $\!\times\!$ 1.4 $\!\times\!$ 1.4 mm & 1.3 $\!\times\!$ 1.3 $\!\times\!$ 4.0 mm  \\
\hdashline
TI, TE/TR & 270-310 ms, 2.3/5.4 ms & 300 ms, 2.3/5.4 ms & 280 ms, 2.1/5.3 ms  & 280 ms, 2.1/5.3 ms  \\
\hdashline
Pre-scan  & N/A & \textless 7 days & \textless 7 days & \textless 2 days \\
\hdashline
Post-scan & 3-27 months & 3-6 months & 30 days & 3-6 months \\
\hline
\end{tabular} }\\
\end{table*}

% Gold standard and evaluation
\textit{MICCAI2018 Atrial Segmentation Challenge} provided LA manual segmentation for the training data, and scars of all LGE MRIs were manually delineated by a well-trained expert.
These manual segmentations were considered as the gold standard.
For LA segmentation evaluation, Dice volume overlap (Dice$_\text{LA}$), average surface distance (ASD) and HD were applied.
For scar quantification evaluation, the manual and (semi-) automatic segmentation results were first projected onto the manually segmented LA surface.
Then, the \textit{Accuracy} measurement of the two areas in the projected surface, Dice of scars (Dice$_\text{s}$) and generalized Dice score (Dice$_\text{g}$) were used as indicators of the accuracy of scar quantification. 
Dice$_\text{s}$ only evaluates one label, while Dice$_\text{g}$ is a weighted Dice score by evaluating the segmentation of all labels \citep{journal/tmi/CrumCH2006,journal/jhe/Zhuang2013}. They are formulated as follows,

\begin{equation}
\begin{array}{l@{\ }l}
 \mathrm{Dice_s} &= \frac {2\left|{S}_{k}^{\textit{auto}} \cap {S}_{k}^{\textit{manual}}\right|} {\left|S_{k}^{\textit{auto}}\right| + \left|S_{k}^{\textit{manual}}\right|},
\end{array}
\label{eq:Dice_s}
\end{equation}

\begin{equation}
\begin{array}{l@{\ }l}
 \mathrm{Dice_g} &= \frac {2\sum_{k=0}^{N_{k}-1}\left| {S}_{k}^{\textit{auto}} \cap {S}_{k}^{\textit{manual}}\right|} {\sum_{k=0}^{N_{k}-1}(\left| S_{k}^{\textit{auto}}\right| + \left|S_{k}^{\textit{manual}}\right|)},
\end{array}
\end{equation}
where $S_{k}^{\textit{auto}}$ and $S_{k}^{\textit{manual}}$ indicate the segmentation results of label $k$ from the automatic method and manual delineation, respectively, and $N_{k}$ is the number of labels. 
In this work, $k$ refers to scars in \Zxhrefeq{eq:Dice_s}, and $N_{k}=2$ (scars and normal walls).

\begin{table*} [t] \center
    \caption{                                                                                                     
    Summary of the quantitative evaluation results of LA segmentation and scar quantification.
    Here, LA$_\text{M}$ denotes that scar quantification is based on the manually segmented LA, while LA$_{\mathrm{U\mbox{-}Net}}$ indicates that it is based on the U-Net$_\mathrm{LA}$-BCE segmentation;
    U-Net$_\mathrm{LA/scar}$ is the separated LA/ scar segmentation directly based on the U-Net architecture with different loss functions.
    AJSQnet refers to the proposed multi-task network architecture, which can achieve joint atrial segmentation and scar quantification. 
    The inter-observer variation (Inter-Ob) is calculated from randomly selected twelve subjects.
     }
\label{tb:result:LAscar}
{\small
\begin{tabular}{  l| l l l | l l l *{7}{@{\ \,} l }}\hline
\multirow{2}*{Method}& \multicolumn{3}{c|}{LA} & \multicolumn{3}{c}{Scar}\\
\cline{2-7}
% \cline{2-3} \cline{5-7}
~      & Dice$_\text{LA}$  &ASD (mm) & HD (mm) & \textit{Accuracy} & Dice$_\text{s}$ &  Dice$_\text{g}$\\
\hline
LA$_\text{M}$+Otsu       &N/A& N/A& N/A &$ 0.750 \pm 0.219 $&  $ 0.420 \pm 0.106 $ &  $ 0.750 \pm 0.188 $\\
LA$_\text{M}$+MGMM       &N/A& N/A& N/A &$ 0.717 \pm 0.250 $&  $ 0.499 \pm 0.148 $ & $ 0.725 \pm 0.239 $\\
LA$_\text{M}$+LearnGC    &N/A& N/A& N/A &$ 0.868 \pm 0.024 $& $ 0.481 \pm 0.151 $ & $ 0.856 \pm 0.029 $\\ %$LA_\text{M} + MS\mbox{-}CNN^0$
\hline
LA$_{\text{U\mbox{-}Net}}$+Otsu  &N/A& N/A& N/A &$ 0.604 \pm 0.339 $& $ 0.359 \pm 0.106 $ & $ 0.567 \pm 0.359 $ \\
LA$_{\text{U\mbox{-}Net}}$+MGMM  &N/A& N/A& N/A &$ 0.579 \pm 0.334 $& $ 0.430 \pm 0.174 $ & $ 0.556 \pm 0.370 $ \\
\hline
U-Net$_\text{LA/scar}$-BCE          &$ 0.889 \pm 0.035 $&  $ 2.12 \pm 0.797 $&  $ 36.4 \pm 23.6 $ &$ 0.866 \pm 0.032 $&  $ 0.357 \pm 0.199 $ & $ 0.843 \pm 0.043 $\\
U-Net$_\text{LA/scar}$-Dice         &$ 0.891 \pm 0.049 $&  $ 2.14 \pm 0.888 $&  $ 35.0 \pm 17.7 $ &$ 0.881 \pm 0.030 $& $ 0.374 \pm 0.156 $ & $ 0.854 \pm 0.041 $\\
U-Net$_\text{LA/scar}$-SE           &$ 0.880 \pm 0.058 $&  $ 2.36 \pm 1.49 $&  $ 25.1 \pm 11.9 $ &$ 0.868 \pm 0.026 $& $ 0.485 \pm 0.129 $ & $ 0.863 \pm 0.026 $\\
\hline\hline
AJSQnet-BCE                            &$ 0.890 \pm 0.042 $&  $ 2.11 \pm 1.01 $&  $ 28.5 \pm 14.0 $ &\bm{$ 0.887 \pm 0.023 $}& $ 0.484 \pm 0.099 $ & \bm{$ 0.872 \pm 0.024 $}\\
AJSQnet-SE                             &$ 0.909 \pm 0.033 $&  $ 1.69 \pm 0.688 $&  $ 22.4 \pm 9.80 $ &$ 0.882 \pm 0.026 $& $ 0.518 \pm 0.110 $ & $ 0.871 \pm 0.024 $\\
AJSQnet-SESA                           &\bm{$ 0.913 \pm 0.032 $}&  \bm{$ 1.60 \pm 0.717 $}& \bm{$ 20.0 \pm 9.59 $} &$ 0.867 \pm 0.032 $& \bm{$ 0.543 \pm 0.097 $} & $ 0.868 \pm 0.028 $\\
\hline\hline
Inter-Ob                           &$ 0.894 \pm 0.011 $& $ 1.807 \pm 0.272 $& $ 17.0 \pm 5.50 $ &$ 0.891 \pm 0.017 $& $ 0.580 \pm 0.110 $ & $ 0.888 \pm 0.022 $\\
\hline
\end{tabular} }\\
\end{table*}

% Implementation
AtrialJSQnet was implemented in PyTorch, running on a computer with 1.90 GHz Intel(R) Xeon(R) E5-2620 CPU and an NVIDIA TITAN X GPU.
We used the SGD optimizer to update the network parameters (weight decay=0.0001, momentum=0.9). 
The initial learning rate was set to 0.001 and divided by 10 every 4000 iterations. 
The balancing parameters in Section \ref{method}, were empirically set as follows, $\lambda_{LA}=0.01$, $\lambda_{scar}=10$, $\lambda_{M_{1}}=0.01$ and $\lambda_{M_{2}}=0.001$, where $\lambda_{LA}$ and $\lambda_{M_{2}}$ was multiplied by 1.1 every 200 iterations.
We adaptively updated the balancing parameter $\lambda_{LA}$ and $\lambda_{M_{2}}$ to progressively increase the complexity of the task.
As we did not have validation set, we randomly selected a model once the training loss tended to converge.
The inference of the networks required about 8 seconds to process one test image.

\subsection{Performance of the proposed method} \label{results}

\subsubsection{Accuracy of LA segmentation and scar quantification} \label{result_accuracy}
\Zxhreftb{tb:result:LAscar} summarizes the quantitative evaluation results of the proposed AJSQnet-SESA.
The average Dice$_\text{LA}$ is $ 0.913 \pm 0.032 $, and the average Dice$_\text{s}$ is $ 0.543 \pm 0.097 $.
To provide a reference for the quantitative evaluation metrics, we conducted a study of inter-observation variation from two manual delineations of LA and scars.
We randomly selected twelve cases from the available data, and two experts manually labeled the scars separately.
As the MICCAI2018 LA challenge did not offer the inter-observer variation value, one expert also generated the manual LA segmentation for the twelve cases.
Note that the two manual segmented scars were projected onto the LA$_\text{M}$ surface for evaluation.
The Dice$_\text{LA}$, ASD and HD of inter-observer variation for LA segmentation were $ 0.894 \pm 0.011 $, $ 1.807 \pm 0.272 $ and $ 17.0 \pm 5.50 $, respectively.
The \textit{Accuracy}, Dice$_\text{s}$ and Dice$_\text{g}$ of inter-observer variation for scar quantification were $ 0.891 \pm 0.017 $, $ 0.580 \pm 0.110 $ and $ 0.888 \pm 0.022 $, respectively.
The accuracy of LA segmentation is comparable to its inter-observer variation, while that of scar quantification is slightly worse than its inter-observer variation.

\begin{figure}[t]\center
	\includegraphics[width=0.48\textwidth]{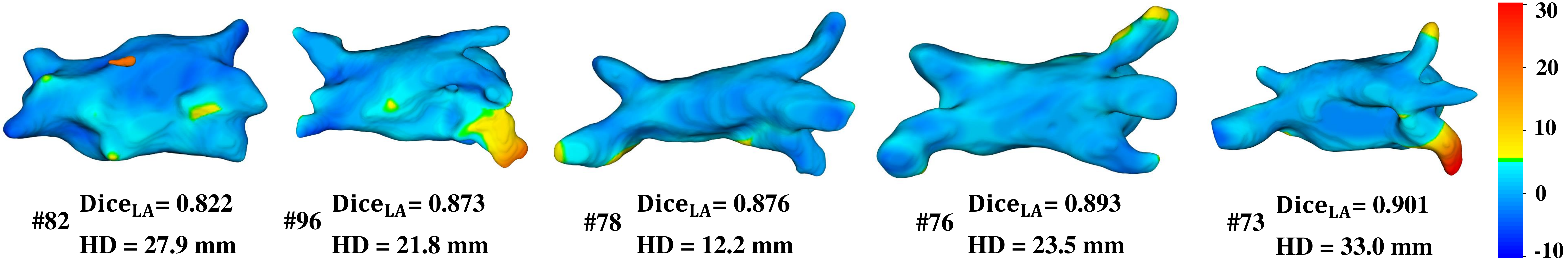}\\[-2ex]
	\caption{3D visualization of the five worst LA segmentation results by the proposed method. The color-coding in the surface refers to the signed surface distance from LA$_\text{auto}$ to LA$_\text{M}$.
	}
	\label{fig:result:3d_results_LA}
\end{figure}

\begin{figure}[t]\center
 \includegraphics[width=0.48\textwidth]{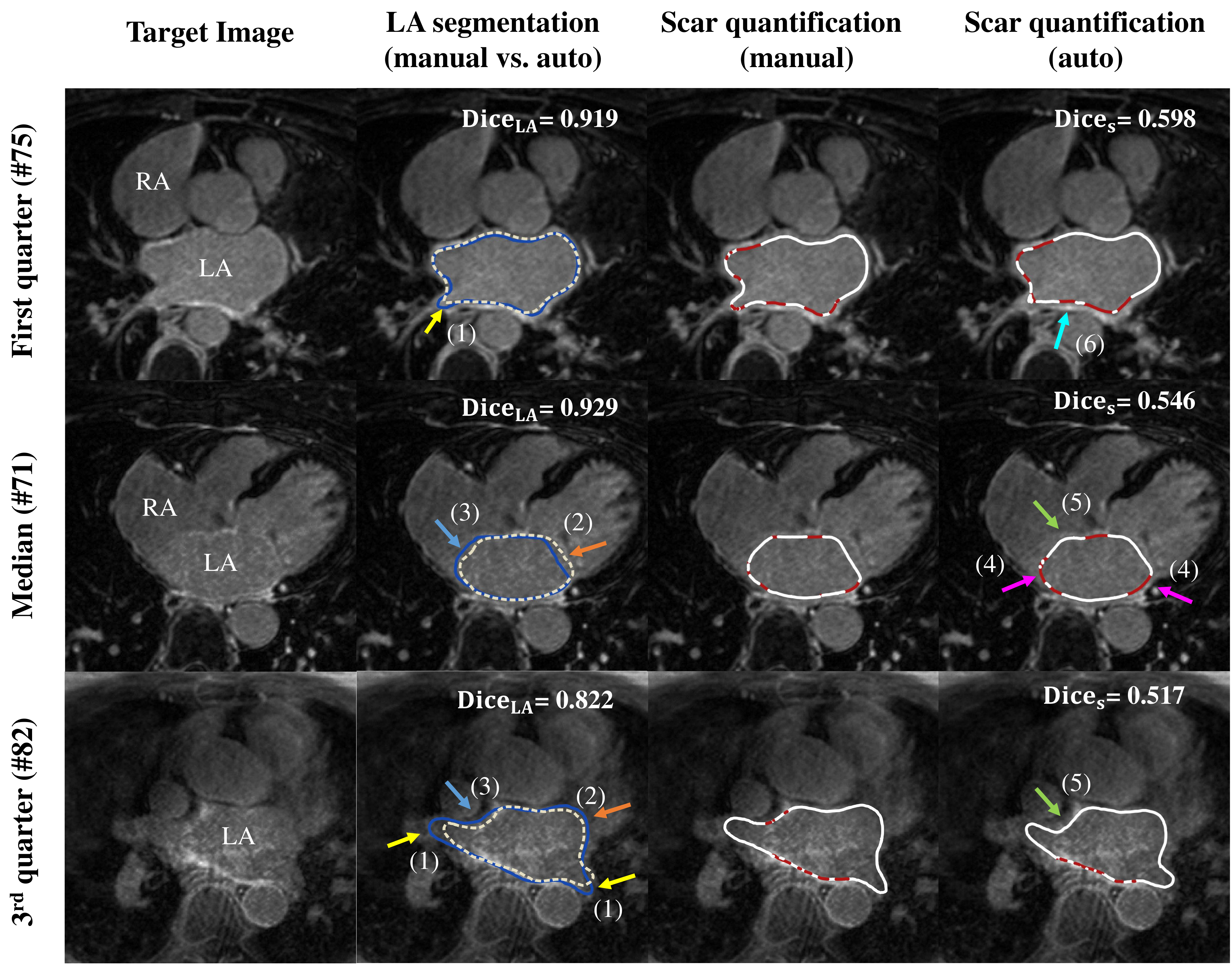}\\[-2ex]
   \caption{
   Axial view of the images, manually and automatically segmented LA, manual scar quantification, and the automatic scar quantification results by the proposed method. 
   In the second column, LA$_\text{M}$ is labeled using \sout{yellow} blue, while LA$_\text{auto}$ is labeled using milky white.
   In the third and fourth columns, the scar and normal regions were separately labeled in red and white. 
   For LA segmentation, arrows (1-3) indicate the misclassification of PV, mitral valve and boundary between LA and RA, respectively.
   For scar quantification, arrow (4) shows that the classification errors of the proposed method due to the blurry boundary;
   arrow (5) shows the regions that were slightly over segmented;
   arrow (6) demonstrates that the proposed method can be robust to inaccurate LA segmentation for scar quantification.} 
\label{fig:result:2dvisual}
\end{figure}

\Zxhreffig{fig:result:3d_results_LA} visualizes five LA segmentation results to illustrate the surface distance between LA$_\text{auto}$ and LA$_\text{M}$.
The five cases were selected from the test set with the worst performance in terms of Dice$_\text{LA}$ by the proposed method.
One can see that although there is a large morphological variation of LA among patients, the proposed method still achieved promising performance.
The obtained LA shape is generally smooth except for case \#82, thanks to the proposed shape regularization term.
The failure of case \#82 could be mainly due to the existence of artifacts in the image, see \Zxhreffig{fig:result:2dvisual}.
The main error of the segmentation is on the pulmonary vein (PV) region, due to its various shapes and complex anatomy.
Besides, some errors are located inside the LA cavity, maybe due to the existence of scars that have complex intensity distributions.
% \Leicolor{}{Note that, compared to many previous works that including additional MRI sequence to assist LGE MRI segmentation, we directly segment LA from LGE MRI.}

\Zxhreffig{fig:result:2dvisual} provides 2D illustrations of LA segmentation and scar quantification results in the axial view from three examples.
The three cases were the first quarter, median and third quarter cases from the test set in terms of Dice$_\text{s}$ by the proposed method.
This illustrates that the method could provide promising performance for segmenting LA, localizing and quantifying atrial scars.
For LA segmentation, we highlight the errors, particularly due to the various and complex PV shape, pointed out by arrow (1).
The errors also occurred in mitral valve and boundary regions between LA and right atrium (RA) (see arrow (2) and (3)), which is mainly due to the poor quality of images.
For scar quantification, there were some slightly over-segmented regions, as arrow (4) points out.
Another type of error occurred in the boundary areas between LA and RA (see arrow (5)), where some scars are hard to detect even for experts.
However, even with LA segmentation errors, the proposed method still could identify the scars at the corresponding location of the projected surface, indicated by arrow (6).
This is mainly attributed to the SE loss, which preserves the spatial information along the perpendicular direction of the surface.

\subsubsection{Model generalization study} \label{result_DG}

\begin{figure*}[t]\center
 \includegraphics[width=0.75\textwidth]{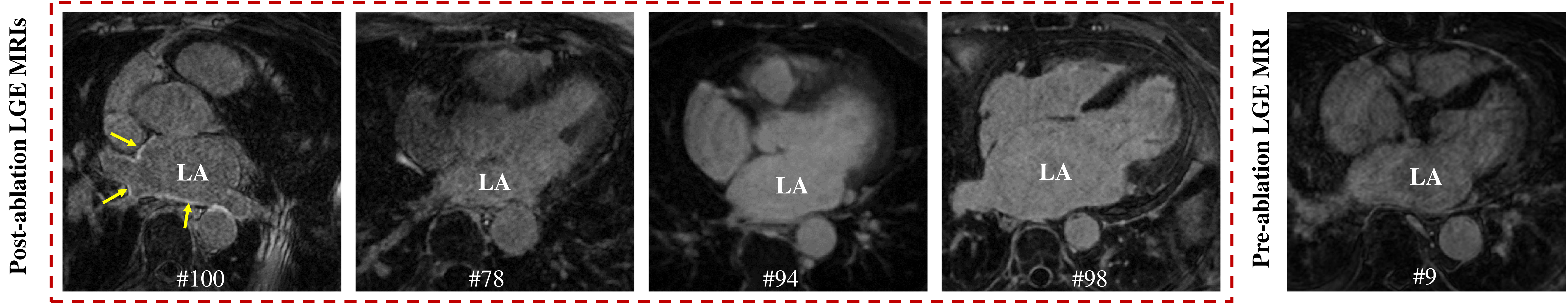}\\[-2ex]
   \caption{Examples of post- and pre-ablation LGE MRI showing the data mismatch problem. The arrows in case \#100 indicate the scarring regions which are clearly enhanced. However, the other three post-ablation LGE MRIs which do not have distinguish scars, are similar to pre-ablation LGE MRIs, such as case \#9.}
\label{fig:result:image}
\end{figure*}

\begin{table*} [t] \center
    \caption{                                                                                               
    Summary of the quantitative evaluation results on different data sources. 
    Additional \textit{ISBI2012 Left Atrium Fibrosis and Scar Segmentation Challenge} dataset (notated using $^{\P}$) was acquired from three centers, i.e., University of Utah (Utah), Beth Israel Deaconess Medical Center (BIDMC), and King’s College London (KCL).
     }
\label{tb:result:DG}
{\footnotesize%\small
%\begin{tabular}{p{2cm}p{2.6cm}|p{1.74cm}p{1.7cm}p{1.7cm}|p{1.74cm}p{1.74cm}p{1.74cm}} 
\begin{tabular}{  l l| l l l | l l l *{8}{@{\ \,} l }}
\hline
\multicolumn{2}{c|}{Data source} & \multicolumn{3}{c|}{LA} & \multicolumn{3}{c}{Scar}\\
\hline
Training data & Test data & Dice$_\text{LA}$  &ASD (mm) & HD (mm) & \textit{Accuracy} & Dice$_\text{s}$ &  Dice$_\text{g}$\\
\hline
40 $I^{\text{post}}$ & 20 $I^{\text{post}}$ &$ 0.913 \pm 0.032 $&  $ 1.60 \pm 0.717 $& $ 20.0 \pm 9.59 $ &$ 0.867 \pm 0.032 $& $ 0.543 \pm 0.097 $ & $ 0.868 \pm 0.028 $\\
\hline
\multirow{3}*{40 $I^{\text{post}}$+20 $I^{\text{pre}}$} & 20 $I^{\text{post}}$ &$ 0.919 \pm 0.024 $&  $ 1.44 \pm 0.428 $& $ 20.3 \pm 9.74 $ &$ 0.875 \pm 0.034 $& $ 0.555 \pm 0.112 $ & $ 0.874 \pm 0.029 $\\
~ & 20 $I^{\text{pre}}$ &$ 0.911 \pm 0.021 $&  $ 1.62 \pm 0.517 $& $ 21.2 \pm 9.77 $ &$ 0.892 \pm 0.031 $& $ 0.405 \pm 0.114 $ & $ 0.885 \pm 0.031 $\\
~ & 20 $I^{\text{post}}$+20 $I^{\text{pre}}$ &$ 0.915 \pm 0.023 $&  $ 1.53 \pm 0.477 $& $ 20.8 \pm 9.64 $ &$ 0.883 \pm 0.033 $& $ 0.480 \pm 0.135 $ & $ 0.880 \pm 0.030 $\\
\hline \hline
\multirow{2}*{40 $I^{\text{post}}$} & 20 $I^{\text{pre}}$ &$ 0.906 \pm 0.021 $&  $ 1.71 \pm 0.598 $& $ 22.0 \pm 9.26 $ &$ 0.875 \pm 0.040 $& $ 0.365 \pm 0.112 $ & $ 0.872 \pm 0.036 $\\
 ~ & 30 $I^{\text{post}\P}$ &$ 0.599 \pm 0.195 $&  $ 6.87 \pm 5.68 $& $ 36.1 \pm 10.5 $ &$ 0.898 \pm 0.044 $& $ 0.218 \pm 0.133 $ & $ 0.928 \pm 0.030 $\\
\hdashline
\multirow{3}*{40 $I^{\text{post}}$} & 10 $I^{\text{post}\P}$ (Utah) &$ 0.650 \pm 0.174 $&  $ 6.37 \pm 1.94 $& $ 39.5 \pm 10.7 $ &$ 0.880 \pm 0.062 $& $ 0.225 \pm 0.116 $ & $ 0.915 \pm 0.043 $\\
~ & 10 $I^{\text{post}\P}$ (BIDMC) &$ 0.648 \pm 0.061 $&  $ 5.42 \pm 1.07 $& $ 33.7 \pm 7.57 $ &$ 0.909 \pm 0.026 $& $ 0.334 \pm 0.050 $ & $ 0.930 \pm 0.019 $\\
~ & 10 $I^{\text{post}\P}$ (KCL) &$ 0.499 \pm 0.267 $&  $ 8.82 \pm 9.60 $& $ 35.2 \pm 13.0 $ &$ 0.905 \pm 0.035 $& $ 0.094 \pm 0.095 $ & $ 0.938 \pm 0.020 $\\
\hline
\end{tabular} }\\
\end{table*}

In addition to the reasons mentioned above, data mismatch can also be a major reason for the bad performance in some cases.
\zxhreffig{fig:result:image} provides four post-ablation LGE MRIs selected from the test data, where the right three cases had the worst performance in terms of Dice$_\text{s}$.
One can see that the image quality of case \#78 is poor with a blurry boundary between LA and its surrounding substructures.
Note that the scars are hard to distinguish even for experts in the three cases.
In contrast, most of LGE MRIs in this dataset resemble case \#100, where enhanced scars are relatively easy to be distinguished visually.
This situation indicates the data mismatch among different subjects, which makes it challenging to segment scars especially with limited training data. 
%It also could be a reason why the performance of the proposed method on scar quantification was worse than its inter-observation.
Note that if we exclude the three special cases, the average Dice$_\text{s}$ is $ 0.577 \pm 0.054 $, which is comparable to its inter-observer value ($ 0.580 \pm 0.110 $) with lower standard deviation.

To verify the mentioned data mismatch problem, we applied the AtrialJSQnet model trained on post-ablation data to a test dataset from the same center, consisting of 20 pre-ablation LGE MRIs. 
Compared to the results on the 20 post-ablation LGE MRIs, the performance was statistically significant worse in terms of Dice$_\text{s}$ ($ 0.543 \pm 0.097 $ vs. $ 0.365 \pm 0.112 $,  $p \textless 0.001$), as presented in \Zxhreftb{tb:result:DG}. 
Previous works showed that the LA scar segmentation and quantification of pre-ablation LGE MRI is more challenging than that of post-ablation LGE MRI \citep{journal/jcmr/Karim2013,journal/MP/yang2018}. 
This is reasonable as the scars on pre-ablation LGE MRIs generally appear more diffusely.
Our experimental results also support this conclusion when we combined pre- and post-ablation LGE MRIs for training and test. 
Therefore, there could be two reasons resulting in the performance decrease: (1) inherent quantification difficulties of pre-ablation LGE MRIs; (2) domain shift between pre- and post-ablation LGE MRIs.
When observing the results on 20 $I^{\text{pre}}$, one can see that the performance was considerably worse in terms of Dice$_\text{s}$ ($ 0.365 \pm 0.112 $ vs. $ 0.405 \pm 0.114 $,  $p = 0.008$) when training data is 40 $I^{\text{post}}$ instead of 40 $I^{\text{post}}$+20 $I^{\text{pre}}$.
However, there was no evident performance decline on the LA segmentation.
Therefore, we conclude that there may exists a distribution shift in the enhancement between pre- and post-ablation LGE MRI from the same center.
%It revealed that pre- and post-ablation LGE MRI from the same center may exist distribution shift in enhancement.

To explore the domain shift between the dataset from different centers, we tested the trained model on another public dataset (notated using $^{\P}$) from \textit{ISBI2012 LA Fibrosis and Scar Segmentation Challenge} \citep{link/LAScarSeg2012}.
As reported in \Zxhreftb{tb:result:DG}, the performance of 30 $I^{\text{post}\P}$ was statistically significant worse ($p  \textless 0.01$) compared to that of 20 $I^{\text{post}}$ for both LA segmentation and scar quantification.
It reveals the limited generalization ability of the trained model on unknown domains from the new centers.
Moreover, the performances among different centers were different, which may indicate the different levels of domain shift between the target domains and source domain.
It is interesting that the performance of $I^{\text{post}\P}$ (Utah) was also poor, although the target domain $I^{\text{post}\P}$ and the source domain $I^{\text{post}}$ were both acquired from Utah.
This could be attribute to the large variation of LGE MRIs from Utah, which were made public in 2012 and 2018, separately.

 \begin{figure*}[t]\center
    % \subfigure[] {\includegraphics[width=0.31\textwidth]{fig_LA_Percentage}} %0.31
    \subfigure[] {\includegraphics[width=0.31\textwidth]{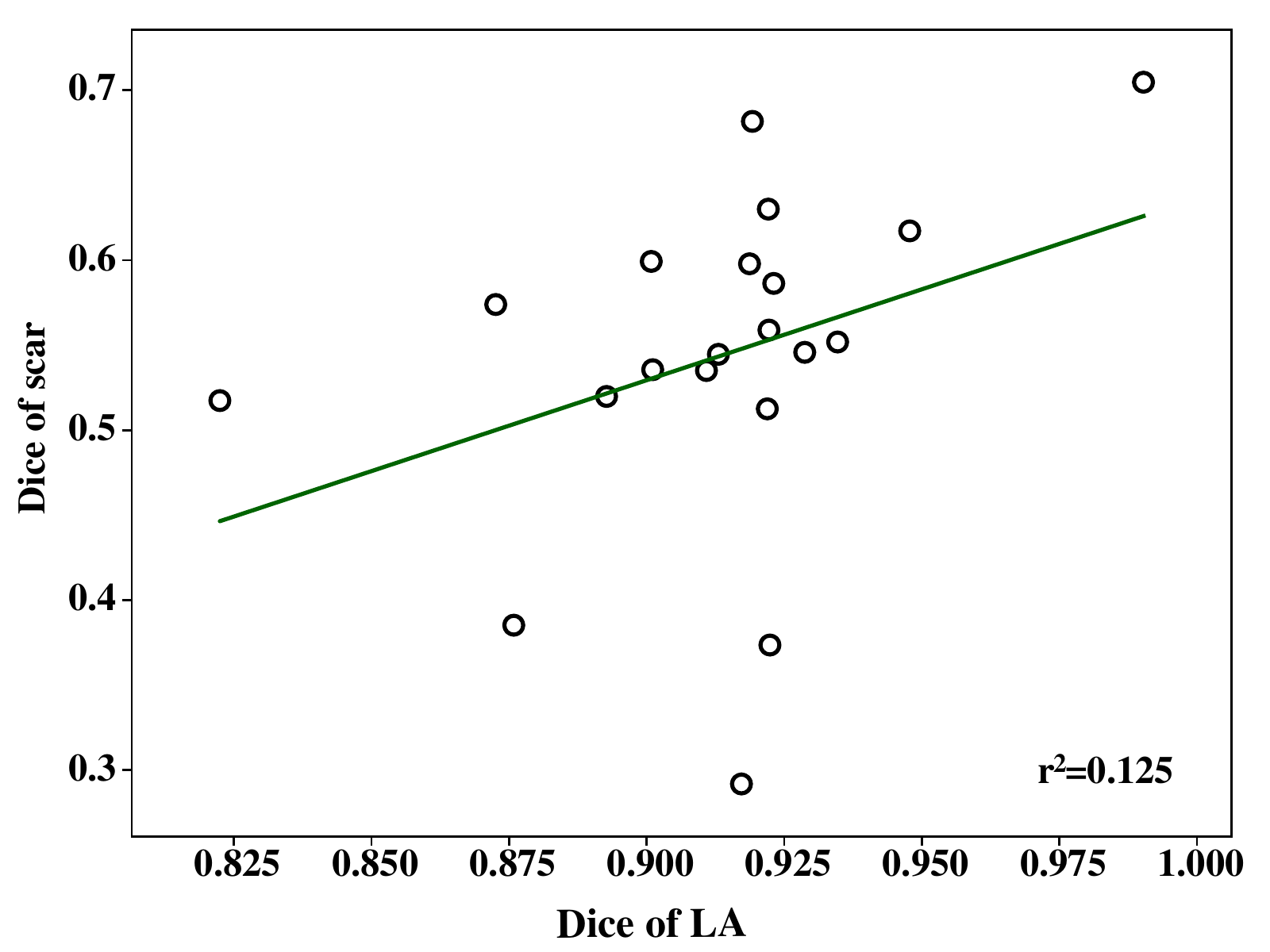}}
    \subfigure[] {\includegraphics[width=0.31\textwidth]{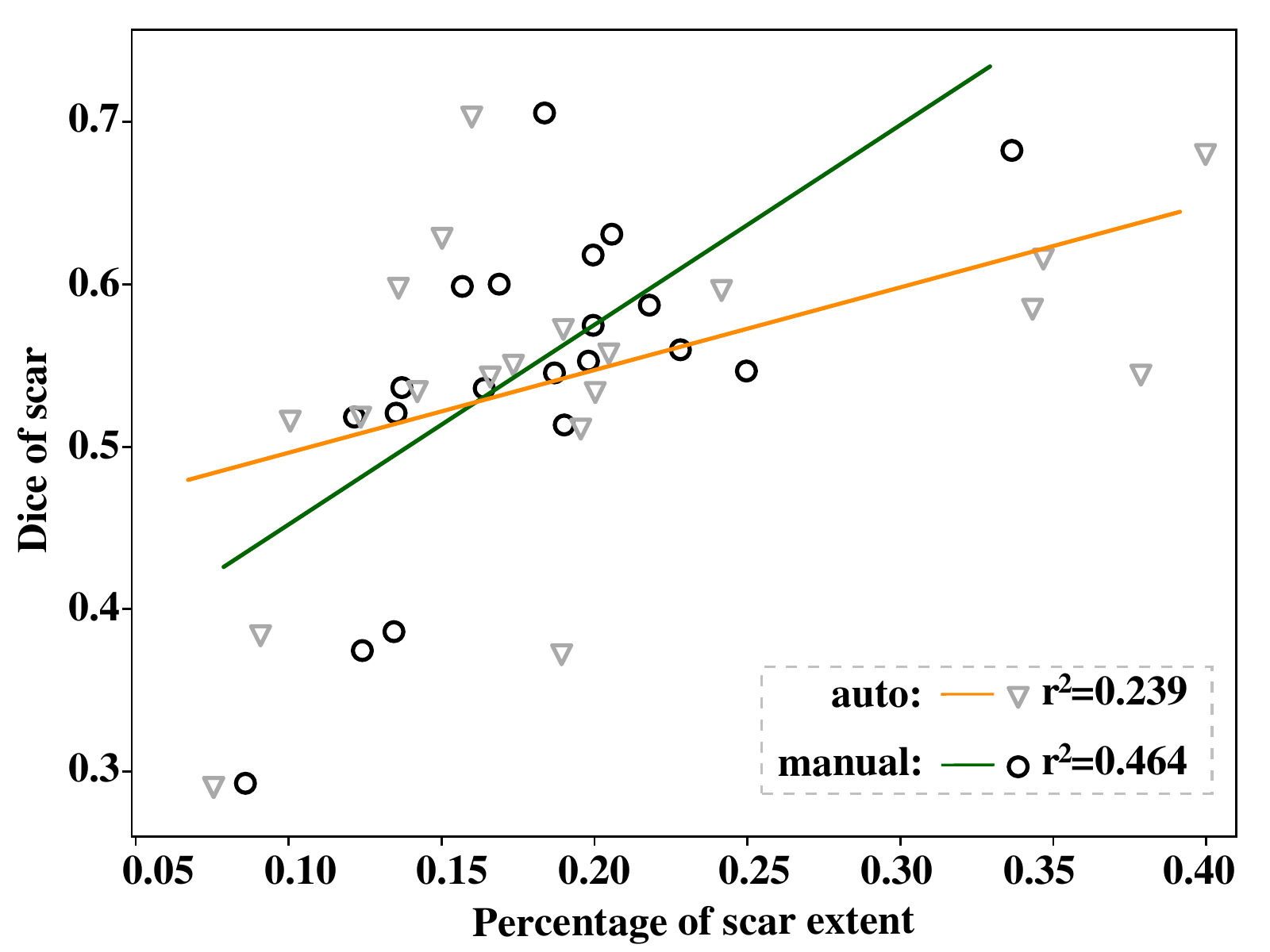}}
    \subfigure[] {\includegraphics[width=0.31\textwidth]{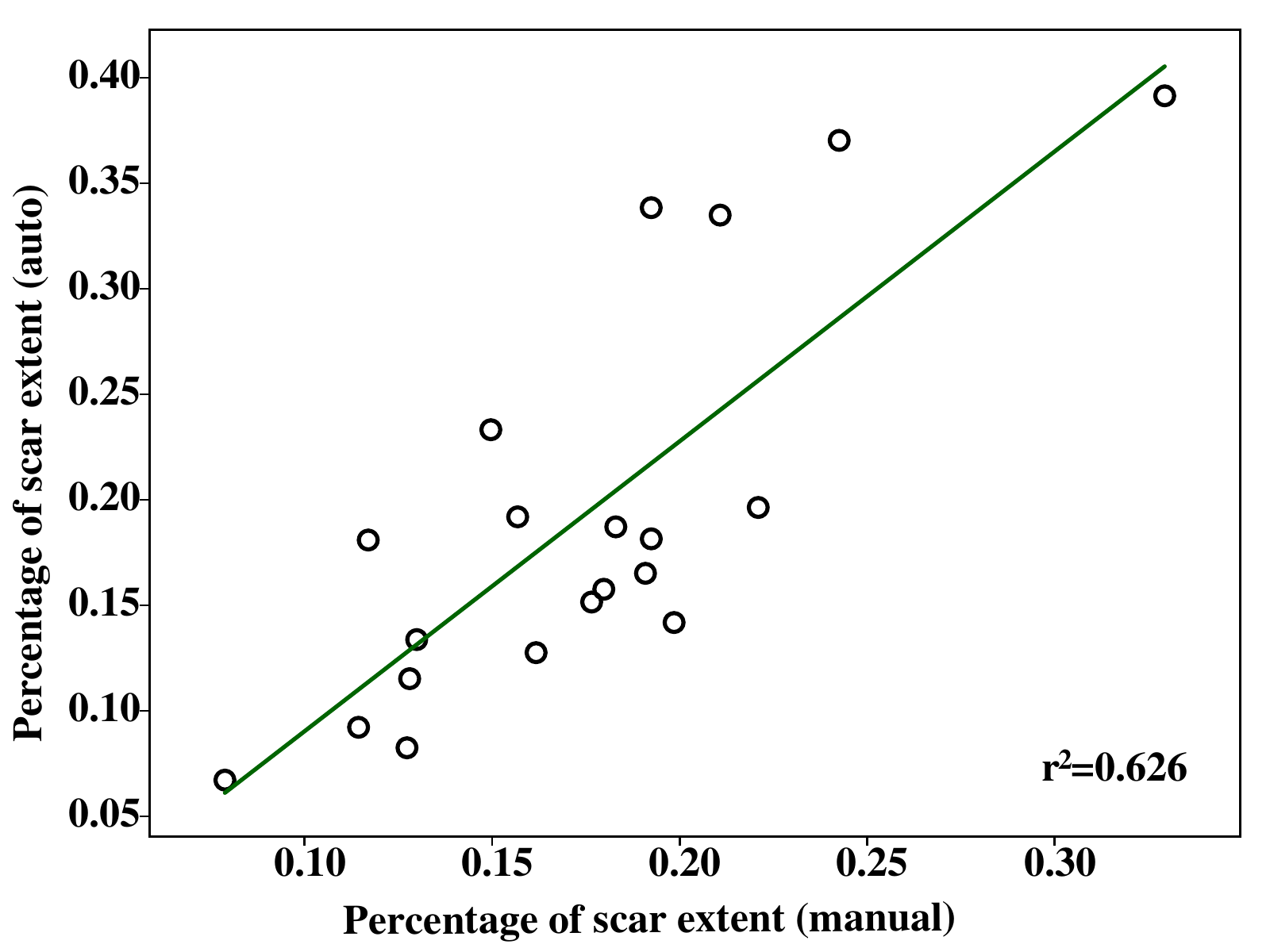}}
	\caption{
		Scatter plots and lines of best fit depicting the correlations between: 
% 		(a) the LA segmentation performance and the scar percentage from the manual (labeled in green) and automatic (in orange) scar segmentation;
		(a) Dice of scars and LA obtained by the proposed method;
		(b) Dice of scars and scar percentages in the manual (labeled in green) and automatic (in orange) scar segmentation;
	    (c) scar percentages from manual scar segmentation and its from automatic scar segmentation.
		}
	\label{fig:result:scatter_plot}
\end{figure*}

\subsubsection{Correlation study} \label{result_correlation}
To further explore the effects related to the scar quantification accuracy, we first analyzed the relationship between Dice$_\text{s}$ and Dice$_\text{LA}$ obtained by the proposed method for each test case.
\Zxhreffig{fig:result:scatter_plot} (a) shows the corresponding scatter plot.
One can see that there is a positive correlation between Dice$_\text{s}$ and Dice$_\text{LA}$, but the correlation is very weak (r$^2=0.125$).
It is reasonable as a more accurate LA segmentation could offer a better initialization for scar quantification, but the proposed method also tends to eliminate the effect of inaccurate LA segmentation. 
Similarly, we analyzed the relation between Dice$_\text{s}$ and scar percentages from both manual and automatic segmentation, as \Zxhreffig{fig:result:scatter_plot} (b) shows.
Here, scar percentage refers to the proportion of scar to the whole LA surface.
The result also shows a positive linear correlation between them.
This indicates that a target with fewer scars presents more difficulties in obtaining a high value of Dice$_\text{s}$.
% \Leicolor{}{This indicates different levels of challenges for different targets, namely, target with less or smaller scars presents more difficulties.}
Besides, one can see that the correlation between Dice$_\text{s}$ and scar percentages from automatic scar segmentation was slightly weaker than that from manual segmentation.
This could be mainly attributed to the misalignment between manual and automatic scar segmentation, which resulted in the difference of scar percentage.
To present the difference, we plotted the manual versus automatic scar percentage for each test subject as two-dimension scatter points in \Zxhreffig{fig:result:scatter_plot} (c).
This relationship quantifies the amount of overlap between manual and automatically extracted scarring regions.
One can see that there is a linear relation between them with a small error (r$^2=0.626$), which demonstrates that the manual and automatic scarring regions generally overlap.
To conclude, the correlation study illustrates a positive but low correlation between the scar quantification accuracy and the LA segmentation accuracy by the proposed method.
It implies that the scar quantification by the proposed method does not rely on accurate LA segmentation results.
Compared to the accuracy of LA segmentation, the accuracy of scar quantification is more related to the scar percentage of the target.
Although the percentage can be slightly different in manual and automatically obtained scar segmentation, the scar distributions in the two segmentations are still generally similar (please refer to \Zxhreffig{fig:result:3d_results_scars}). 

\subsubsection{Parameter studies} \label{exp:param_study}

\begin{figure*}[!t]\center
	\subfigure[] {\includegraphics[width=0.47\textwidth]{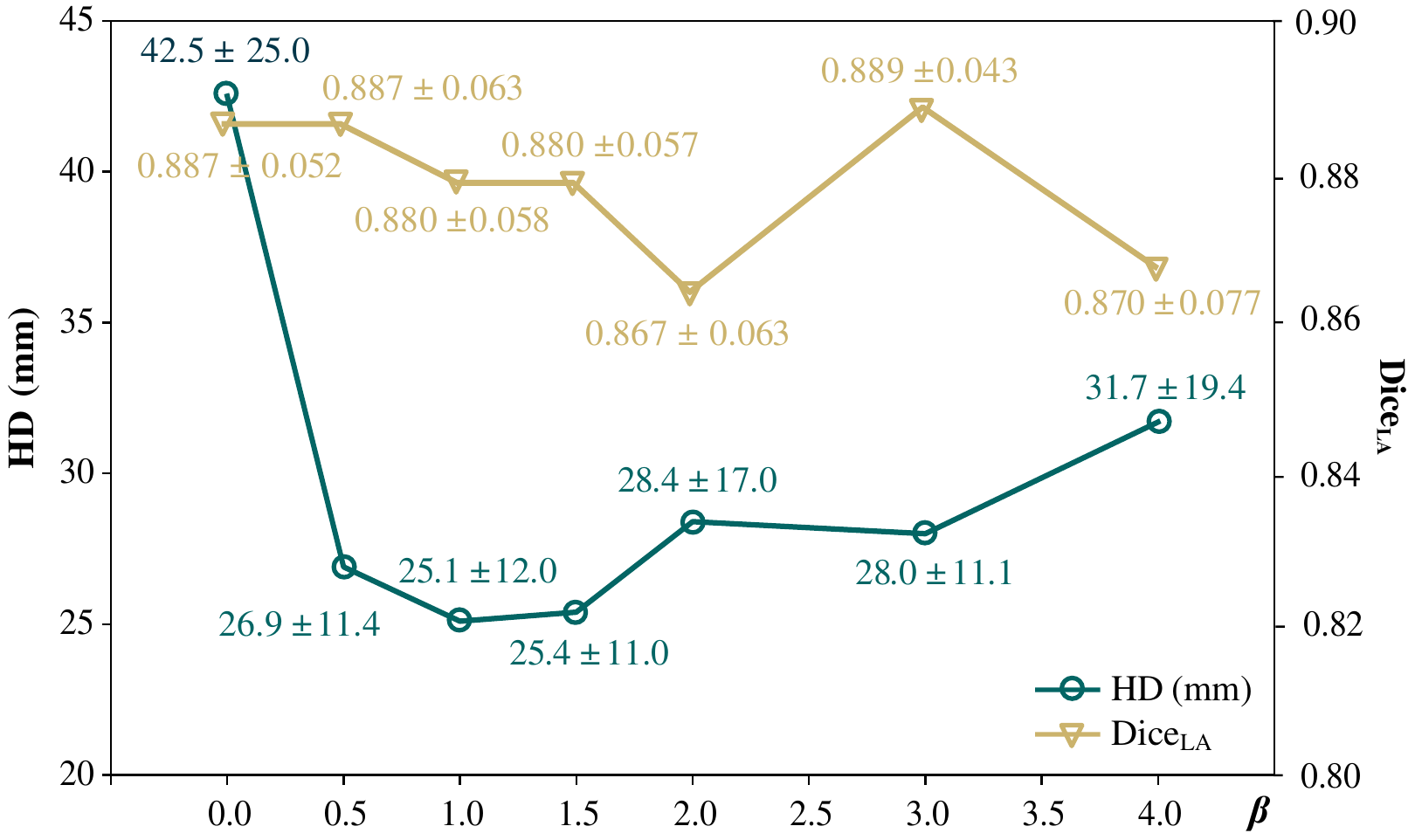}}
	\subfigure[] {\includegraphics[width=0.44\textwidth]{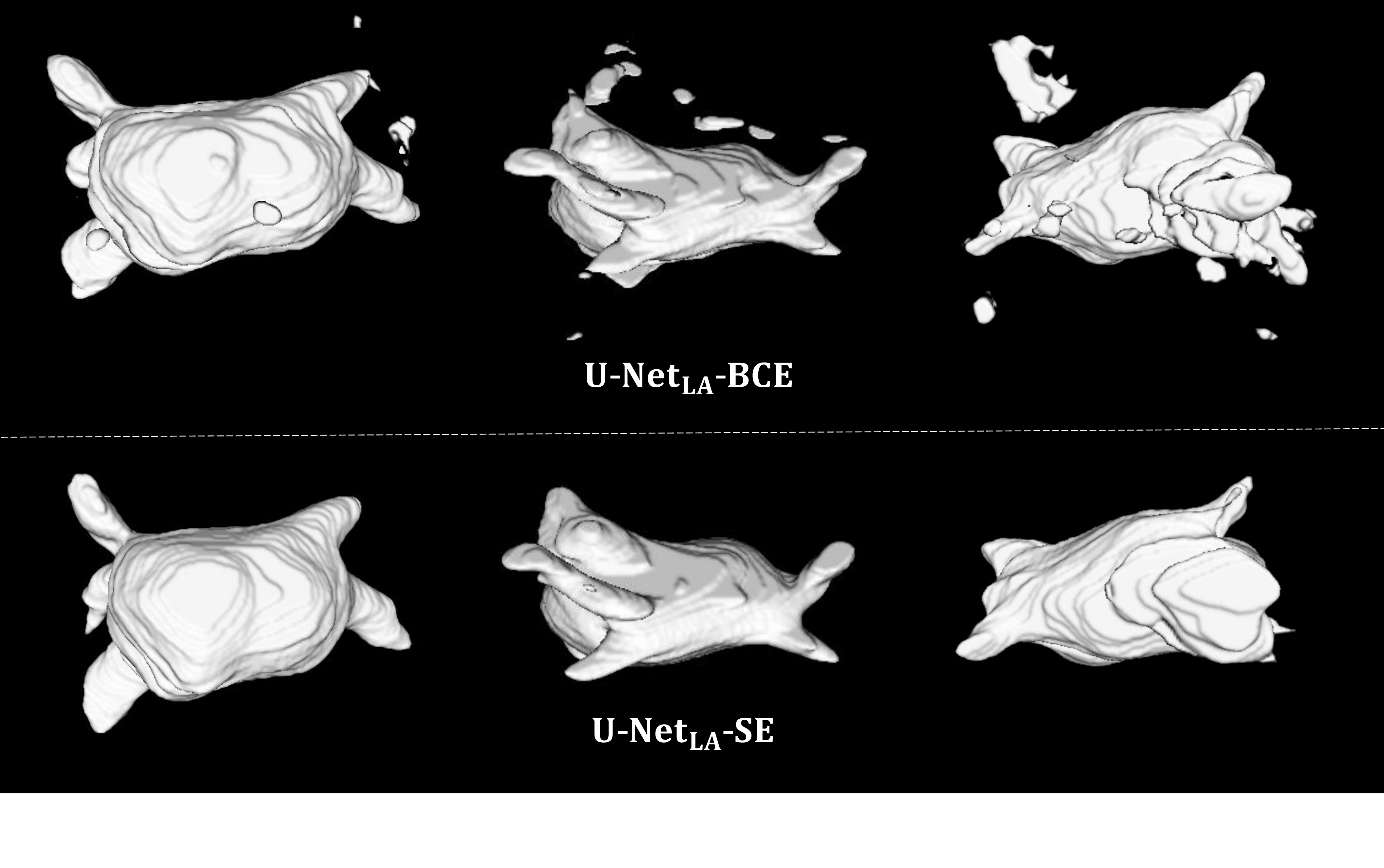}}
	\caption{
		Quantitative and qualitative evaluation results of the proposed SE loss for LA segmentation: (a) Dice and HD of the LA segmentation results after combining the SE loss, i.e., U-Net$_\text{LA}$-SE with different $\beta$ for DTM (see \Zxhrefeq{eq:SDM}); 
		(b) 3D visualization of the LA segmentation results of four typical cases by U-Net$_\text{LA}$-BCE and U-Net$_\text{LA}$-SE ($\beta=1$).}
	\label{fig:result:LA_SE}
\end{figure*}

To explore the effectiveness of the SE loss, we compared the results of the proposed scheme for LA segmentation using different values of $\beta$ for DTM (see \Zxhrefeq{eq:SDM}).
\Zxhreffig{fig:result:LA_SE} (a) provides the results in terms of Dice$_\text{LA}$ and HD.
The introduction of the SE loss significantly improved the HD of the resulting segmentation ($p<0.001$), although the Dice$_\text{LA}$ may not be very different.
Also, U-Net$_\text{LA}$-SE showed stable performance with different values of $\beta$ except for too extreme values.
% \zxhcolor{}{This is reasonable, as ...}
In this work, we used $\beta=1$.

To evaluate the effect of DPM and $\mathcal L_{scar}^{SE}$ definitions, in \Zxhrefeq{eq:SE_scar}, for the proposed method, we tested different options and combinations.
The results are presented in \Zxhreftb{tb:result:param}.
One can see that the combination ($exp^{-|\phi(x)|}$ + L2) obtained the best results in terms of Dice$_\text{s}$.
We also explored another function for the conversion of the DTM to a probability map, namely $expit(-\phi(x))$.
Compared to $exp^{-|\phi(x)|}$, $expit(-\phi(x))$ converges slower towards 0 when voxels are far away from the target.
One can see that using $expit$ to construct DPMs obtained similar results to the proposed scheme without statistically significant difference for the majority of the evaluation measurements.  
We further normalized the DPMs of scars and normal wall by considering the probability of the background, and compared it with unnormalized maps.
Both maps (unnormalized and normalized) had a similar effect for scar quantification (Dice$_\text{s}$: $ 0.543 \pm 0.097 $ vs. $ 0.523 \pm 0.118 $, $p=0.086$; $ 0.535 \pm 0.116 $ vs. $ 0.525 \pm 0.111 $, $p=0.127$).
There are various metrics to estimate the \textit{difference} between two probability distributions, such as Kullback-Leibler divergence (KL-divergence) \citep{journal/AMS/Kullback1951}, Hellinger distance \citep{journal/JM/hellinger1909} and L2 norm \citep{journal/NC/Sugiyama2013}.
Compared to KL-divergence, Hellinger distance and L2 norm are more effective in practice due to their high computational efficiency and numerical stability \citep{journal/JCSE/Sugiyama2013}.  
We therefore compared the performance of the proposed algorithm calculating $\mathcal L_{scar}^{SE}$ via L2 and Hellinger distance, but did not find any statistically significant differences for all evaluation measurements except for the \textit{Accuracy} of scars.
% Hellinger distance, a metric to measure the difference between two probability distributions, can be regarded as the probabilistic analog of Euclidean distance.
These experimental results showed that the proposed method is generally robust to different definitions of DPM and $\mathcal L_{scar}^{SE}$. 
Note that we did not convert the DTMs using the exponential function to construct $\mathcal L_{LA}^{SE}$, since this is not appropriate for reducing outliers in the LA segmentation. 
For instance, if one uses $exp^{-x}$ for the conversion of $\phi(x)$, $exp^{-\phi(x)}$ would be a positive value.
Thus, penalties would be assigned to all voxels regardless of their correct or incorrect classification, which is not plausible for the LA segmentation.
% It may attribute that we obtain final scar quantification results by comparing the DPMs of scars and normal wall, which enables the DPM and $\mathcal L_{scar}^{SE}$ to be more robust to different definitions. 

\begin{table*} [t] \center
    \caption{
    Performance of the proposed method based on different definitions of DPM and $\mathcal L_{scar}^{SE}$.
    Here, $expit(x)= 1/(1+exp^{-x})$, and Dagger ($^\dagger$) refers to the methods adopting the normalized probabilities by considering the probabilities of scars, normal wall and background.
    Note that $exp^{(-|\phi(x)|)\dagger}$ equals to $softmax(-|\phi(x)|)$.
    Asterisks ($^*$: $p \textless 0.01$, $^{**}$: $p \textless 0.001$) indicate that the methods obtained statistically significant different results compared to the proposed combination, i.e., $exp^{-|\phi(x)|}$ + L2.
     } 
\label{tb:result:param}
{\footnotesize%\small
\begin{tabular}{p{1.6cm}p{2.4cm}|p{1.75cm}p{1.75cm}p{1.75cm}|p{1.89cm}p{1.8cm}p{1.9cm}} 
%\begin{tabular}{  l l| l l l | l l l *{8}{@{\ \,} l }}
\hline
\multicolumn{2}{c|}{Definitions} & \multicolumn{3}{c|}{LA} & \multicolumn{3}{c}{Scar}\\
\hline
DPM & $\mathcal L_{scar}^{SE}$ & Dice$_\text{LA}$  &ASD (mm) & HD (mm) & \textit{Accuracy} & Dice$_\text{s}$ &  Dice$_\text{g}$\\
\hline
$exp^{-|\phi(x)|}$ & L2 &$ 0.913 \pm 0.032 $&  $ 1.60 \pm 0.717 $&  $ 20.0 \pm 9.59 $ &$ 0.867 \pm 0.032 $& $ 0.543 \pm 0.097 $ & $ 0.868 \pm 0.028 $\\
\hline
$expit(-\phi(x))$  & L2 &$ 0.913 \pm 0.029$&  $ 1.60 \pm 0.570$&  $ 21.4 \pm 9.33$ & $ 0.880 \pm 0.030^{**} $ & $ 0.535 \pm 0.116$ & $ 0.874 \pm 0.028$\\ 
$exp^{-|\phi(x)|}$ & Hellinger distance &$ 0.913 \pm 0.029$&  $ 1.57 \pm 0.546$&  $ 22.6 \pm 13.5$ &$ 0.876 \pm 0.033^{*} $ & $ 0.522 \pm 0.121$ & $ 0.871 \pm 0.031$\\
%$expit(-\phi(x))$  & Hellinger distance &$ 0.911 \pm 0.031$&  $ 1.60 \pm 0.581$&  $ 19.7 \pm 9.69$ &$ 0.884 \pm 0.029^{**} $ & $ 0.522 \pm 0.121$ &$ 0.875 \pm 0.028^\S $\\
\hdashline
$exp^{(-|\phi(x)|)\dagger}$  & L2 &$ 0.909 \pm 0.030$&  $ 1.65 \pm 0.590$&  $ 20.6 \pm 9.79$ & $ 0.879 \pm 0.030^{*}$ & $ 0.523 \pm 0.118 $ & $ 0.873 \pm 0.027 $\\
$expit(-\phi(x))^\dagger$  & L2 &$ 0.909 \pm 0.031$&  $ 1.64 \pm 0.573$&  $ 20.0 \pm 10.3$ & $ 0.865 \pm 0.037$ & $ 0.525 \pm 0.111 $ & $ 0.865 \pm 0.033$\\ 
%$expit(-\phi(x))^\dagger$  & Hellinger distance &$ 0.910 \pm 0.030$&  $ 1.65 \pm 0.568$&  $ 21.6 \pm 9.47 $ & $ 0.882 \pm 0.030 $ & $ 0.520 \pm 0.116 $ & $ 0.874 \pm 0.028 $\\ 
\hline
\end{tabular} }\\
\end{table*}

\subsection{Ablation study and comparisons with literature} \label{ablation_study}
\subsubsection{Ablation study} \label{exp_ablation}
We performed an ablation study by comparing the results of U-Net$_\text{LA/scar}$-BCE, U-Net$_\text{LA/scar}$-SE, AJSQnet-BCE, AJSQnet-SE, and the proposed method AJSQnet-SESA. 
Here, U-Net$_\text{LA/scar}$ denotes the original U-Net architecture \citep{conf/MICCAI/ronneberger2015} for LA segmentation or scar quantification.
AJSQnet indicates that the methods are based on the architecture in \zxhreffig{fig:method:network}.
BCE, Dice, SE, SA and SESA refer to the different loss functions.
\Zxhreftb{tb:result:LAscar} presents the quantitative results for LA segmentation and scar quantification.

For LA segmentation, combining the proposed SE loss performed better than only using the BCE loss based on both U-Net and AJSQnet.
\Zxhreffig{fig:result:LA_SE} (b) visualizes three examples for illustrating the difference in the results with or without using the SE loss.
One can see that with the SE loss, U-Net$_\text{LA}$-SE evidently reduced clutter and disconnected parts in the prediction compared to U-Net$_\text{LA}$-BCE.

For scar quantification, the SE loss also showed promising performance compared to the conventional losses in terms of Dice$_\text{s}$.
The AJSQnet based methods (AJSQnet-BCE and AJSQnet-SE), performed better than the corresponding U-Net based methods (U-Net$_\text{LA/scar}$-BCE and U-Net$_\text{LA/scar}$-SE). 
Therefore, the LA segmentation and scar quantification both benefited from the proposed joint optimization scheme compared to achieving the two tasks separately.
The results of scar quantification were further improved after introducing the newly-designed SA loss in terms of Dice$_\text{s}$ ($p\leq0.001$), but with a slightly worse \textit{Accuracy} ($p\leq0.001$) and Dice$_\text{g}$ ($p>0.1$) compared to AJSQnet-BCE. 
It may be due to the fact that AJSQnet-SESA tends to slightly over-segment scars compared to AJSQnet-BCE, which in turn tends to achieve under-segmentation, as \Zxhreffig{fig:result:3d_results_scars} illustrated. 
However, AJSQnet-SESA had much higher \textit{Sensitivity} (0.558 vs. 0.380) but lower \textit{Specificity} (0.915 vs. 0.971) compared to AJSQnet-BCE.
% Sensitivity measures the proportion of correctly identified scars, and specificity measures the proportion of correctly segmented normal wall regions. 
% \zxhcolor{}{Note that in clinics, we prefer to identify/ segment as much as possible all the scars, i.e., high sensitivity, though it may marginally sacrifice specificity by minor over-segmentation.---lei: I doubt this proposal. In the paper of Marta "Mind the gap", they aim to find the missed region that is supposed to be scars. Anyway, we need to double-check this.}
%However, the proposed method could obtain a smoother result compared to AJSQnet-BCE, as shown in \Zxhreffig{fig:result:3d_results_scars}. 

\begin{table} [t] \center
    \caption{The results of LA segmentation without using and using different DTM-based shape constrain methods. 
     }
\label{tb:result:LA}
{\small
% \begin{tabular}{l| lllll}
\begin{tabular}{p{26mm}|p{17.4mm}|p{16.2mm}|p{14.2mm}}
\hline
Method & Dice$_\text{LA}$  &ASD (mm) & HD (mm)\\
\hline
U-Net$_\text{LA}$-BCE             &$ 0.892 \pm 0.063 $&  $ 1.99 \pm 0.957 $&  $ 35.2 \pm 17.8 $ \\
\hline
U-Net$_\text{LA}$-SE              &$ 0.893 \pm 0.054 $&  $ 1.92 \pm 1.09 $&  $ 25.7 \pm 13.8 $\\
\citet{journal/MP/dangi2019}      &$ 0.863 \pm 0.120 $&  $ 2.54 \pm 2.47 $&  $ 29.3 \pm 15.0 $\\
\citet{journal/CVIU/audebert2019} &$ 0.882 \pm 0.093 $&  $ 2.14 \pm 1.51 $&  $ 29.1 \pm 15.0 $\\
\citet{conf/xue2019}              &$ 0.883 \pm 0.090 $&  $ 2.22 \pm 1.67 $&  $ 32.0 \pm 15.7 $\\
\hline
\end{tabular} }\\
\end{table}

\subsubsection{Comparison experiment} \label{exp_comparison}
For LA segmentation, we compared our proposal with other three different solutions using DTM for shape constrain \citep{journal/MP/dangi2019,journal/CVIU/audebert2019,conf/xue2019}.
Here, we modified the U-Net$_\text{LA}$ to construct networks with the same parameter setting for the three methods, respectively. %as backbone network architecture
\Zxhreftb{tb:result:LA} presents the quantitative results with and without using DTM for regularization in LA segmentation.
One can see that all these shape constrain strategies were effective to reduce the HD compared to U-Net$_\text{LA}$-BCE.
However, they performed worse in terms of Dice$_\text{LA}$, but without significant differences ($p>0.1$) except for the scheme of \citet{conf/xue2019} ($p=0.004$). 
The proposed scheme U-Net$_\text{LA}$-SE performed better than the three state-of-the-art DTM-based spatial constrain methods, though the difference was not significant for the schemes of \citet{journal/MP/dangi2019} and \citet{journal/CVIU/audebert2019} ($p>0.1$).
Note that compared to the three methods, the proposed scheme does not require the modification of the segmentation network for DTM prediction, but directly employs the ground truth DTM to encode spatial information.

For scar quantification, four state-of-the-art algorithms, i.e., Otsu \citep{journal/tmi/Ravanelli2014}, multi-component GMM (MGMM) \citep{journal/TBME/liu2017}, LearnGC \citep{journal/MedIA/li2020} and U-Net$_\text{scar}$ with different loss functions, were studied for comparison.
\Zxhreftb{tb:result:LAscar} presents the quantitative results.
Here, LA$_\text{M}$ denotes that scar quantification is based on the manually segmented LA, while LA$_{\text{U\mbox{-}Net}}$ indicates that it is based on the segmentation of U-Net$_\text{LA}$-BCE.
One can see that the three (semi-) automatic methods generally obtained acceptable results, but relied on an accurate initialization of LA.
LearnGC had a similar result compared to MGMM in Dice$_\text{s}$ based on LA$_\text{M}$, but its \textit{Accuracy} and Dice$_\text{g}$ were higher.
The proposed method performed significantly better ($p\leq0.001$) than all the fully automatic methods in terms of Dice$_\text{s}$.

\begin{figure}[t]\center
	\includegraphics[width=0.48\textwidth]{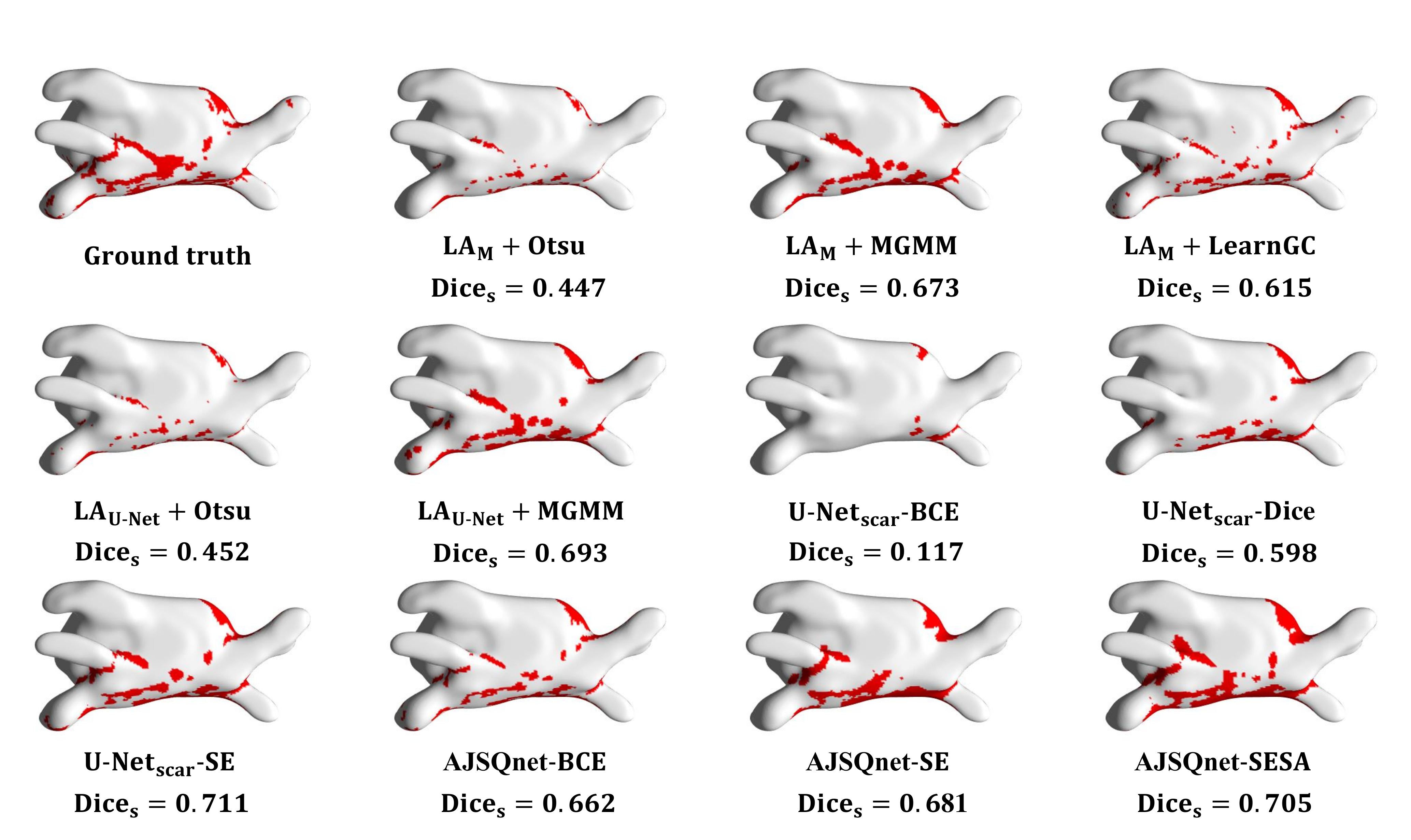}\\[-2ex]
	\caption{3D visualization of the LA scar localization by the eleven methods. The scarring areas are labeled in red on the LA surface, which is constructed from LA$_\text{M}$ labeled in white.
	}
	\label{fig:result:3d_results_scars}
\end{figure}

\Zxhreffig{fig:result:3d_results_scars} illustrates segmentation and quantification results of scars from the mentioned methods in \Zxhreftb{tb:result:LAscar}.
One can see that Otsu and U-Net$_\text{scar}$-BCE tended to under-segment the scars, and the introduction of Dice and SE loss alleviated this problem for U-Net$_\text{scar}$.
Additionally, the SE loss was more effective compared to Dice loss, which was consistent with the quantitative results in \Zxhreftb{tb:result:LAscar}.
Besides employing a new loss, the modification of the network also improved the results when comparing the results of U-Net$_\text{scar}$-BCE and AJSQnet-BCE.
MGMM and LearnGC both detected most of the scars, but LearnGC has the potential advantage of small scar detection, which is one of the main challenges of scar quantification as mentioned in Section \ref{result_correlation}.
Compared to LearnGC, the proposed method could also detect small and discrete scars but with a smoother segmentation result, and achieved an end-to-end scar quantification and projection.
% \Leicolor{}{Note that when the scarring area is small, the Dice score of the scars tends to be lower.}

\begin{figure}[t]\center
 \includegraphics[width=0.48\textwidth]{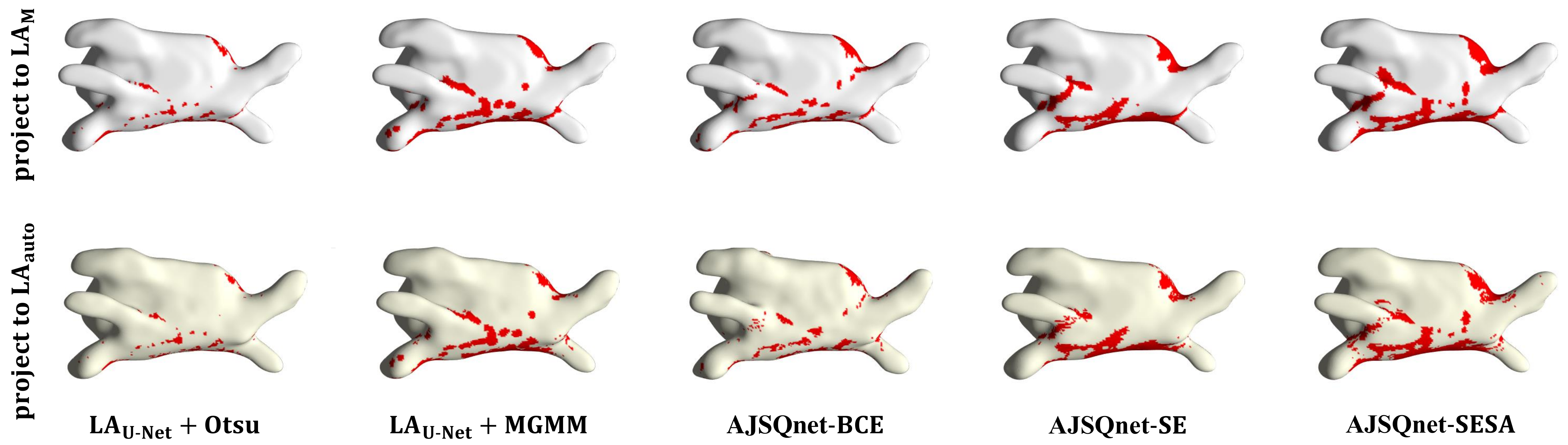}\\[-2ex]
   \caption{The 3D visualization of the LA scar localization results by the five methods with automatic LA segmentation in \Zxhreftb{tb:result:LAscar}. The scars are projected on the LA surface which can be manually delineated LA (LA$_\text{M}$, surface labeled in white) or automatically segmented LA (LA$_\text{auto}$, surface labeled in yellow). 
   }
\label{fig:result:projection}
\end{figure}

%surface projection
To compare the 3D scar quantification results on two different surfaces (LA$_{\text{M}}$ and LA$_{\text{auto}}$), we visualized the results of the five methods with automatic LA segmentation in \Zxhreftb{tb:result:LAscar}, as shown in \Zxhreffig{fig:result:projection}.   
One can see that although scars were projected onto two different surfaces, similar scar patterns can be observed thanks to the accurate LA segmentation.
This is consistent with the quantitative evaluation results, which are also similar when scars were projected onto the different LA surfaces.
For example, the proposed method AJSQnet-SESA obtained $0.542\pm0.105$ and $0.543\pm0.097$ Dice$_\text{s}$ ($p$=0.860), when scars were projected onto manually and automatically segmented LA, respectively.

\subsubsection{Results from the literature} \label{literature results}

For comparison and reference, we summarized the literature results for the LA segmentation and scar quantification from LGE MRI, as presented in \Zxhreftb{tb:table:review}.
For LA segmentation, all these methods reported similar or worse Dice of LA than ours in this work.
For scar quantification, most of the automatic methods among the seven works obtained similar results with ours, except for \citet{journal/MP/yang2018} and \citet{journal/MedIA/li2020}, which yielded better Dice of scars than ours but there was no evident difference in terms of \textit{Accuracy}.
% \citet{conf/MI/perry2012} had the manual LA wall segmentation, and \citet{journal/TEHM/karim2014, journal/tmi/Ravanelli2014} employed semi-automatic approaches for the LA segmentation.
Only two works \citep{conf/MICCAI/chen2018,journal/FGCS/yang2020} simultaneously segmented LA and scars and reported better results than ours, but they did not report inter-observation values.
Note that it can be difficult to conduct a fair cross-study comparison due to the differences in datasets, manual interventions, and evaluation metrics.
For example, \citet{journal/MedIA/li2020} reported a good performance on their dataset but obtained a worse result on the public data compared to our proposed method, as \Zxhreftb{tb:result:LAscar} shows.

\newcommand{\tabincell}[2]{\begin{tabular}{@{}#1@{}}#2\end{tabular}}
\begin{table*} [t] \center
    \caption{
    Overview of previous methods for scar quantification and segmentation in LA. Abbreviations: number of subjects (N); inter-observer variation in terms of Dice (Inter-ob); registration-based method (Reg); Society of Photo-Optical Instrumentation Engineers (SPIE), IEEE Journal of Translational Engineering in Health and Medicine (TEHM), IEEE transactions on medical imaging (TMI), Medical physics (MP), Medical Image Computing and Computer-Assisted Intervention (MICCAI), Medical Image Analysis (MedIA), Future Generation Computer Systems (FCGS).
     }
\label{tb:table:review}
{\small
\begin{tabular}{ l| l *{5}{@{\ \,} l }}\hline
Work     &  N &  Method (LA) & Method (scar) & Dice (LA) &  Dice (scar) & Inter-ob (scar)\\
\hline
\citet{conf/MI/perry2012}, SPIE         &34  & N/A$^\text{manual}$           & K-means            & N/A               & $0.807 \pm 0.106$  & $ 0.786 \pm 0.072 $\\
\citet{journal/TEHM/karim2014}, TEHM    &15  & Reg$^\text{semi-auto}$        & GMM + Graph-cuts   & N/A               & $>0.8$             &  N/A \\
\citet{journal/tmi/Ravanelli2014}, TMI  &10  & Otsu + Reg$^\text{semi-auto}$ & NVI                & $0.90$            & $0.850 \pm 0.070$  &  N/A \\
                                        &10 &  Otsu + Reg$^\text{auto}$      & NVI                & $0.60$            & $0.600 \pm 0.210$  &  N/A \\
\citet{journal/MP/yang2018}, MP         &37  & MAS$^\text{auto}$             & Super-pixels + SVM & $0.89$            & $0.790 \pm 0.050$  & N/A \\                      
\citet{conf/miccai/wu2018}, MICCAI      &36  & MAS$^\text{auto}$             & MvMM               & N/A               & $0.556 \pm 0.187$  & N/A \\
\citet{journal/MedIA/li2020}, MedIA     &58  & MAS$^\text{auto}$             & LearnGC            & $0.898 \pm 0.044$ & $0.702 \pm 0.071$  &  $ 0.695 \pm 0.049 $ \\
\hline\hline
\citet{conf/MICCAI/chen2018}, MICCAI    &100 & MvTT$^\text{auto}$            & MvTT               & $0.908 \pm 0.031$ & $0.776 \pm 0.146$  & N/A \\
\citet{journal/FGCS/yang2020}, FCGS     &190 & MvTT$^\text{auto}$            & MvTT               & $0.931 \pm 0.019$ & $0.870$            & N/A \\
\hline
\end{tabular} }
\end{table*}

\section{Discussion and conclusion} \label{discussion}

In this work, we have proposed an end-to-end framework for joint LA segmentation, scar projection and scar quantification, which incorporates spatial and shape information.
Two major methodological contributions have been introduced.
One is the adoption of an alternative ground truth representation in the form of DTM, which is embedded into the network by the SE loss.
The SE loss forces the network to assign different weights to each voxel, to learn the spatial information of the target.
It effectively removes outliers and reduces the overall HD for LA segmentation, as shown in \Zxhreffig{fig:result:LA_SE}.
More importantly, it does not rely on any prior and does not require any network modification, so it can be a promising alternative for other complex shape regularization methods \citep{conf/MICCAI/yue2019,journal/MedIA/li2020,journal/MedAI/kamnitsas2017,conf/MICCAI/zeng2019}.
The second contribution is the optimization of an end-to-end AtrialJSQnet for simultaneous LA segmentation and scar quantification with a surface projection.
The spatial relationship of LA and scars is explicitly learned by employing a shape attention scheme, i.e., SA loss, which helps to project the scars onto the LA surface at the same time.
The surface projection avoids the difficulty of providing an accurate and demanding LA wall segmentation,
and the SE loss further mitigates the effect of inaccurate LA segmentation, as demonstrated in Section \ref{results}.
The combination of SE and SA loss also alleviates the class-imbalance problem in the scar quantification compared to Dice and BCE loss.
Besides, both LA segmentation and scar quantification benefit from the joint optimization scheme compared to achieving them separately (see Section \ref{exp_ablation}).
We employed sixty images with manual delineation for experiments, and the proposed AJSQnet-SESA method demonstrates better performance compared to the conventional approaches (see Section \ref{exp_comparison}).
The mean Dice$_\text{LA}$ and HD for LA segmentation are 0.913 and 20.0 mm, respectively, while \textit{Accuracy} and Dice$_\text{s}$ for quantifying LA scars are 0.867 and 0.543, respectively.
The results are comparable to those of inter-observer variation of LA (Dice$_\text{LA}$=0.894, HD=17.0) and slightly worse than that of scars (\textit{Accuracy}=0.891, Dice$_\text{s}$=0.580).

\paragraph{Limitation and remaining challenges}
There are two main challenges for the LA segmentation.
The first one is the processing of the high variability of atrial shapes, including global features such as size, and local features such as the number, position and orientation of the PVs, as shown in \zxhreffig{fig:result:3d_results_LA}.
Another challenge is from the poor quality and complex intensity distribution of LGE MRI (see \zxhreffig{fig:result:2dvisual} and \zxhreffig{fig:result:image}).
Combining other MRI sequences from the same subject with LGE MRI by registration, is a common strategy to tackle this challenge, as discussed above.
In this work, we proposed to use AtrialJSQnet to directly extract the features of LGE MRI and learn the spatial information of LA.
% \zxhcolor{}{The simultaneous prediction of LA and scars and the proposed shape attention scheme, also made a \textbf{constrain} for LA segmentation.}

For the scar quantification, the first challenge is to distinguish artifacts from the boundary regions, as we discussed in Section \ref{result_accuracy} and showed in \zxhreffig{fig:result:2dvisual}.
Providing accurate LA walls is typically used to solve this problem \citep{journal/jcmr/Karim2013,conf/MI/perry2012}.
Here, we propose to use spatial encoding and shape attention to learn the spatial information of scars around the wall.
However, the misclassification caused by this issue could still occur due to the limited training data.
Another challenge is due to the various intensity distributions of scars (see \zxhreffig{fig:result:image}), resulting in the data mismatch, further increasing the difficulty of training.

A major limitation of this work is the lack of model generalization ability, even for the dataset collected from the same center.
This is mainly due to the lack of a standardized LA LGE MRI acquisition protocol.
For example, there is no consensus on the option and dose of the contrast agent, to the best of our knowledge, nor on the timing of image acquisition after contrast agent administration \citep{journal/MedIA/li2021}. 
Though \citet{journal/JCMR/kramer2020} proposed to standardize disease-specific protocols to clinical MRI by recommending acquisition protocols for LGE MRI, but it is primarily for myocardial infarction rather than AF.
\citet{journal/LRM/craft2021} offered a more elaborated supplement to the standardized guidelines \citep{journal/JCMR/kramer2020} for the LA LGE imaging in AF.
However, the feasibility of standardizing the LA LGE MRI acquisition protocol to improve model generalization ability is yet to be verified. 
Note that even for well-established cine MRI acquisition techniques, there still exist severe domain shift problems between multi-center, multi-vendor, and multi-disease images for ventricle segmentation \citep{journal/TMI/campello2021}. 
Therefore, it is desired to develop LA LGE MRI computing models with efficient generalization abilities for multi-center and multi-vendor data.
Current LA LGE MRI computing approaches are mainly evaluated on center- and vendor-specific LGE MRI.
Although \textit{Left Atrium Fibrosis and Scar Segmentation Challenge} offered multi-center and multi-scanner data, the model generalization abilities of the benchmark algorithms were not explored \citep{journal/jcmr/Karim2013}.
Recently, we performed a preliminary investigation on the domain shift issue among multi-center LGE MRIs \citep{conf/MICCAI/li2021} . 
The results revealed that the performance of commonly used segmentation models degraded dramatically on the unknown domain for the LA segmentation, and domain generalization strategies have showed potential in mitigating the performance decay. 
We concluded that there exists significant scope for the algorithmic exploitation in terms of improving the generalization capability of the model for the LA LGE MRI computing.

% The exploration of the above challenges/ limitations leads us to future directions.
% First, the model generalization ability should be improved to process out-of-distribution samples due to data mismatch.
% We need to further investigate the domain shift between pre- and post-ablation LGE MRI from the same center, as well as the labeling variations in LGE MRI from different centers.
% Second, more multi-center and multi-vendor AF data will be collected, labeled and ranked according to its image quality.
% The final goal is to develop an unified model for the LA segmentation and scar quantification with the ability to effectively generalize to multi-center and multi-vendor data.

% %Conclusion
% In conclusion, the proposed method has the potential to locate and quantify the scars, and it outperforms conventional methods.
% It is fully automatic and can generate reliable and accurate results, comparable to the inter-observer variation from two experts.
% Hence, it can be useful in the clinical care of AF patients.

\section*{Acknowledgment}
This work was funded by the National Natural Science Foundation of China (grant no. 61971142, 62111530195 and 62011540404) and the development fund for Shanghai talents (no. 2020015), and L. Li was partially supported by the CSC Scholarship.
JA Schnabel and VA Zimmer would like to acknowledge funding from a Wellcome Trust IEH Award (WT 102431), an EPSRC program Grant (EP/P001009/1), and the Wellcome/EPSRC Center for Medical Engineering (WT 203148/Z/16/Z).
We thank Prof. Jichao Zhao and Dr. Zhaohan Xiong for providing the LA dataset, Prof. Oscar Camara for useful discussions and Miss Xin Wen for helping with the data processing.

\bibliographystyle{model2-names}%elsarticle-harv}%elsarticle-num}
\bibliography{A_ref}

\begin{thebibliography}{59}
\expandafter\ifx\csname natexlab\endcsname\relax\def\natexlab#1{#1}\fi
\providecommand{\url}[1]{\texttt{#1}}
\providecommand{\href}[2]{#2}
\providecommand{\path}[1]{#1}
\providecommand{\DOIprefix}{doi:}
\providecommand{\ArXivprefix}{arXiv:}
\providecommand{\URLprefix}{URL: }
\providecommand{\Pubmedprefix}{pmid:}
\providecommand{\doi}[1]{\href{http://dx.doi.org/#1}{\path{#1}}}
\providecommand{\Pubmed}[1]{\href{pmid:#1}{\path{#1}}}
\providecommand{\bibinfo}[2]{#2}
\ifx\xfnm\relax \def\xfnm[#1]{\unskip,\space#1}\fi
%Type = Article
\bibitem[{Audebert et~al.(2019)Audebert, Boulch, Le~Saux and
  Lef{\`e}vre}]{journal/CVIU/audebert2019}
\bibinfo{author}{Audebert, N.}, \bibinfo{author}{Boulch, A.},
  \bibinfo{author}{Le~Saux, B.}, \bibinfo{author}{Lef{\`e}vre, S.},
  \bibinfo{year}{2019}.
\newblock \bibinfo{title}{Distance transform regression for spatially-aware
  deep semantic segmentation}.
\newblock \bibinfo{journal}{Computer Vision and Image Understanding}
  \bibinfo{volume}{189}, \bibinfo{pages}{102809}.
%Type = Article
\bibitem[{Avendi et~al.(2016)Avendi, Kheradvar and
  Jafarkhani}]{journal/MedAI/avendi2016}
\bibinfo{author}{Avendi, M.}, \bibinfo{author}{Kheradvar, A.},
  \bibinfo{author}{Jafarkhani, H.}, \bibinfo{year}{2016}.
\newblock \bibinfo{title}{A combined deep-learning and deformable-model
  approach to fully automatic segmentation of the left ventricle in cardiac
  {MRI}}.
\newblock \bibinfo{journal}{Medical image analysis} \bibinfo{volume}{30},
  \bibinfo{pages}{108--119}.
%Type = Article
\bibitem[{Breu and Gil(1995)}]{journal/PAMI/1995}
\bibinfo{author}{Breu, H.}, \bibinfo{author}{Gil, J.}, \bibinfo{year}{1995}.
\newblock \bibinfo{title}{Linear time euclidean distance transform algorithms}.
\newblock \bibinfo{journal}{IEEE Transactions on Pattern Analysis \& Machine
  Intelligence} \bibinfo{volume}{17}, \bibinfo{pages}{529--533}.
%Type = Article
\bibitem[{Campello et~al.(2021)Campello, Gkontra, Izquierdo, Mart{\'\i}n-Isla,
  Sojoudi, Full, Maier-Hein, Zhang, He, Ma et~al.}]{journal/TMI/campello2021}
\bibinfo{author}{Campello, V.M.}, \bibinfo{author}{Gkontra, P.},
  \bibinfo{author}{Izquierdo, C.}, \bibinfo{author}{Mart{\'\i}n-Isla, C.},
  \bibinfo{author}{Sojoudi, A.}, \bibinfo{author}{Full, P.M.},
  \bibinfo{author}{Maier-Hein, K.}, \bibinfo{author}{Zhang, Y.},
  \bibinfo{author}{He, Z.}, \bibinfo{author}{Ma, J.}, et~al.,
  \bibinfo{year}{2021}.
\newblock \bibinfo{title}{Multi-centre, multi-vendor and multi-disease cardiac
  segmentation: The {M\&M}s challenge}.
\newblock \bibinfo{journal}{IEEE Transactions on Medical Imaging} .
%Type = Inproceedings
\bibitem[{Chen et~al.(2018a)Chen, Bai and Rueckert}]{conf/STACOM/chen2018}
\bibinfo{author}{Chen, C.}, \bibinfo{author}{Bai, W.},
  \bibinfo{author}{Rueckert, D.}, \bibinfo{year}{2018}a.
\newblock \bibinfo{title}{Multi-task learning for left atrial segmentation on
  {GE-MRI}}, in: \bibinfo{booktitle}{International Workshop on Statistical
  Atlases and Computational Models of the Heart},
  \bibinfo{organization}{Springer}. pp. \bibinfo{pages}{292--301}.
%Type = Inproceedings
\bibitem[{Chen et~al.(2018b)Chen, Yang, Gao, Ni, Angelini, Mohiaddin, Wong,
  Zhang, Du, Zhang et~al.}]{conf/MICCAI/chen2018}
\bibinfo{author}{Chen, J.}, \bibinfo{author}{Yang, G.}, \bibinfo{author}{Gao,
  Z.}, \bibinfo{author}{Ni, H.}, \bibinfo{author}{Angelini, E.},
  \bibinfo{author}{Mohiaddin, R.}, \bibinfo{author}{Wong, T.},
  \bibinfo{author}{Zhang, Y.}, \bibinfo{author}{Du, X.},
  \bibinfo{author}{Zhang, H.}, et~al., \bibinfo{year}{2018}b.
\newblock \bibinfo{title}{Multiview two-task recursive attention model for left
  atrium and atrial scars segmentation}, in: \bibinfo{booktitle}{International
  Conference on Medical Image Computing and Computer-Assisted Intervention},
  \bibinfo{organization}{Springer}. pp. \bibinfo{pages}{455--463}.
%Type = Inproceedings
\bibitem[{Chen et~al.(2019)Chen, Bortsova, Ju{\'a}rez, van Tulder and
  de~Bruijne}]{conf/MICCAI/chen2019}
\bibinfo{author}{Chen, S.}, \bibinfo{author}{Bortsova, G.},
  \bibinfo{author}{Ju{\'a}rez, A.G.U.}, \bibinfo{author}{van Tulder, G.},
  \bibinfo{author}{de~Bruijne, M.}, \bibinfo{year}{2019}.
\newblock \bibinfo{title}{Multi-task attention-based semi-supervised learning
  for medical image segmentation}, in: \bibinfo{booktitle}{International
  Conference on Medical Image Computing and Computer-Assisted Intervention},
  \bibinfo{organization}{Springer}. pp. \bibinfo{pages}{457--465}.
%Type = Article
\bibitem[{Chugh et~al.(2014)Chugh, Havmoeller, Narayanan, Singh, Rienstra,
  Benjamin, Gillum, Kim, McAnulty~Jr, Zheng et~al.}]{journal/cir/chugh2014}
\bibinfo{author}{Chugh, S.S.}, \bibinfo{author}{Havmoeller, R.},
  \bibinfo{author}{Narayanan, K.}, \bibinfo{author}{Singh, D.},
  \bibinfo{author}{Rienstra, M.}, \bibinfo{author}{Benjamin, E.J.},
  \bibinfo{author}{Gillum, R.F.}, \bibinfo{author}{Kim, Y.H.},
  \bibinfo{author}{McAnulty~Jr, J.H.}, \bibinfo{author}{Zheng, Z.J.}, et~al.,
  \bibinfo{year}{2014}.
\newblock \bibinfo{title}{Worldwide epidemiology of atrial fibrillation: a
  global burden of disease 2010 study}.
\newblock \bibinfo{journal}{Circulation} \bibinfo{volume}{129},
  \bibinfo{pages}{837--847}.
%Type = Article
\bibitem[{Craft et~al.(2021)Craft, Li, Bhatti and Cao}]{journal/LRM/craft2021}
\bibinfo{author}{Craft, J.}, \bibinfo{author}{Li, Y.}, \bibinfo{author}{Bhatti,
  S.}, \bibinfo{author}{Cao, J.J.}, \bibinfo{year}{2021}.
\newblock \bibinfo{title}{How to do left atrial late gadolinium enhancement: a
  review}.
\newblock \bibinfo{journal}{La radiologia medica} , \bibinfo{pages}{1--11}.
%Type = Article
\bibitem[{Crum et~al.(2006)Crum, Camara and Hill}]{journal/tmi/CrumCH2006}
\bibinfo{author}{Crum, W.R.}, \bibinfo{author}{Camara, O.},
  \bibinfo{author}{Hill, D.L.}, \bibinfo{year}{2006}.
\newblock \bibinfo{title}{Generalized overlap measures for evaluation and
  validation in medical image analysis}.
\newblock \bibinfo{journal}{IEEE transactions on medical imaging}
  \bibinfo{volume}{25}, \bibinfo{pages}{1451--1461}.
%Type = Article
\bibitem[{Dangi et~al.(2019)Dangi, Linte and Yaniv}]{journal/MP/dangi2019}
\bibinfo{author}{Dangi, S.}, \bibinfo{author}{Linte, C.A.},
  \bibinfo{author}{Yaniv, Z.}, \bibinfo{year}{2019}.
\newblock \bibinfo{title}{A distance map regularized {CNN} for cardiac cine
  {MR} image segmentation}.
\newblock \bibinfo{journal}{Medical physics} \bibinfo{volume}{46},
  \bibinfo{pages}{5637--5651}.
%Type = Article
\bibitem[{Duan et~al.(2019)Duan, Bello, Schlemper, Bai, Dawes, Biffi,
  de~Marvao, Doumoud, O’Regan and Rueckert}]{journal/TMI/duan2019}
\bibinfo{author}{Duan, J.}, \bibinfo{author}{Bello, G.},
  \bibinfo{author}{Schlemper, J.}, \bibinfo{author}{Bai, W.},
  \bibinfo{author}{Dawes, T.J.}, \bibinfo{author}{Biffi, C.},
  \bibinfo{author}{de~Marvao, A.}, \bibinfo{author}{Doumoud, G.},
  \bibinfo{author}{O’Regan, D.P.}, \bibinfo{author}{Rueckert, D.},
  \bibinfo{year}{2019}.
\newblock \bibinfo{title}{Automatic 3{D} bi-ventricular segmentation of cardiac
  images by a shape-refined multi-task deep learning approach}.
\newblock \bibinfo{journal}{IEEE transactions on medical imaging}
  \bibinfo{volume}{38}, \bibinfo{pages}{2151--2164}.
%Type = Article
\bibitem[{Hellinger(1909)}]{journal/JM/hellinger1909}
\bibinfo{author}{Hellinger, E.}, \bibinfo{year}{1909}.
\newblock \bibinfo{title}{Neue begr{\"u}ndung der theorie quadratischer formen
  von unendlichvielen ver{\"a}nderlichen.}
\newblock \bibinfo{journal}{Journal f{\"u}r die reine und angewandte
  Mathematik} \bibinfo{volume}{1909}, \bibinfo{pages}{210--271}.
%Type = Inproceedings
\bibitem[{Ji et~al.(2018)Ji, van~der Geest, Nazarian, Lelieveldt and
  Tao}]{conf/ip/Ji2018}
\bibinfo{author}{Ji, Y.}, \bibinfo{author}{van~der Geest, R.J.},
  \bibinfo{author}{Nazarian, S.}, \bibinfo{author}{Lelieveldt, B.P.},
  \bibinfo{author}{Tao, Q.}, \bibinfo{year}{2018}.
\newblock \bibinfo{title}{Advanced two-layer level set with a soft distance
  constraint for dual surfaces segmentation in medical images}, in:
  \bibinfo{booktitle}{Medical Imaging 2018: Image Processing}, p.
  \bibinfo{pages}{105743B}.
%Type = Article
\bibitem[{Kamnitsas et~al.(2017)Kamnitsas, Ledig, Newcombe, Simpson, Kane,
  Menon, Rueckert and Glocker}]{journal/MedAI/kamnitsas2017}
\bibinfo{author}{Kamnitsas, K.}, \bibinfo{author}{Ledig, C.},
  \bibinfo{author}{Newcombe, V.F.}, \bibinfo{author}{Simpson, J.P.},
  \bibinfo{author}{Kane, A.D.}, \bibinfo{author}{Menon, D.K.},
  \bibinfo{author}{Rueckert, D.}, \bibinfo{author}{Glocker, B.},
  \bibinfo{year}{2017}.
\newblock \bibinfo{title}{Efficient multi-scale 3{D} {CNN} with fully connected
  {CRF} for accurate brain lesion segmentation}.
\newblock \bibinfo{journal}{Medical image analysis} \bibinfo{volume}{36},
  \bibinfo{pages}{61--78}.
%Type = Article
\bibitem[{Karim et~al.(2014)Karim, Arujuna, Housden, Gill, Cliffe, Matharu,
  Gill, Rindaldi, O'Neill, Rueckert et~al.}]{journal/TEHM/karim2014}
\bibinfo{author}{Karim, R.}, \bibinfo{author}{Arujuna, A.},
  \bibinfo{author}{Housden, R.J.}, \bibinfo{author}{Gill, J.},
  \bibinfo{author}{Cliffe, H.}, \bibinfo{author}{Matharu, K.},
  \bibinfo{author}{Gill, J.}, \bibinfo{author}{Rindaldi, C.A.},
  \bibinfo{author}{O'Neill, M.}, \bibinfo{author}{Rueckert, D.}, et~al.,
  \bibinfo{year}{2014}.
\newblock \bibinfo{title}{A method to standardize quantification of left atrial
  scar from delayed-enhancement {MR} images}.
\newblock \bibinfo{journal}{IEEE journal of translational engineering in health
  and medicine} \bibinfo{volume}{2}, \bibinfo{pages}{1--15}.
%Type = Article
\bibitem[{Karim et~al.(2018)Karim, Blake, Inoue, Tao, Jia, Housden, Bhagirath,
  Duval, Varela, Behar et~al.}]{journal/MedAI/karim2018}
\bibinfo{author}{Karim, R.}, \bibinfo{author}{Blake, L.E.},
  \bibinfo{author}{Inoue, J.}, \bibinfo{author}{Tao, Q.}, \bibinfo{author}{Jia,
  S.}, \bibinfo{author}{Housden, R.J.}, \bibinfo{author}{Bhagirath, P.},
  \bibinfo{author}{Duval, J.L.}, \bibinfo{author}{Varela, M.},
  \bibinfo{author}{Behar, J.M.}, et~al., \bibinfo{year}{2018}.
\newblock \bibinfo{title}{Algorithms for left atrial wall segmentation and
  thickness--evaluation on an open-source {CT} and {MRI} image database}.
\newblock \bibinfo{journal}{Medical image analysis} \bibinfo{volume}{50},
  \bibinfo{pages}{36--53}.
%Type = Article
\bibitem[{Karim et~al.(2013)Karim, Housden, Balasubramaniam, Chen, Perry,
  Uddin, Al-Beyatti, Palkhi, Acheampong, Obom et~al.}]{journal/jcmr/Karim2013}
\bibinfo{author}{Karim, R.}, \bibinfo{author}{Housden, R.J.},
  \bibinfo{author}{Balasubramaniam, M.}, \bibinfo{author}{Chen, Z.},
  \bibinfo{author}{Perry, D.}, \bibinfo{author}{Uddin, A.},
  \bibinfo{author}{Al-Beyatti, Y.}, \bibinfo{author}{Palkhi, E.},
  \bibinfo{author}{Acheampong, P.}, \bibinfo{author}{Obom, S.}, et~al.,
  \bibinfo{year}{2013}.
\newblock \bibinfo{title}{Evaluation of current algorithms for segmentation of
  scar tissue from late gadolinium enhancement cardiovascular magnetic
  resonance of the left atrium: an open-access grand challenge}.
\newblock \bibinfo{journal}{Journal of Cardiovascular Magnetic Resonance}
  \bibinfo{volume}{15}, \bibinfo{pages}{105}.
%Type = Article
\bibitem[{Karimi and Salcudean(2019)}]{journal/TMI/karimi2019}
\bibinfo{author}{Karimi, D.}, \bibinfo{author}{Salcudean, S.E.},
  \bibinfo{year}{2019}.
\newblock \bibinfo{title}{Reducing the {H}ausdorff distance in medical image
  segmentation with convolutional neural networks}.
\newblock \bibinfo{journal}{IEEE transactions on medical imaging} .
%Type = Article
\bibitem[{Karimi et~al.(2018)Karimi, Samei, Kesch, Nir and
  Salcudean}]{journal/JCARS/karimi2018}
\bibinfo{author}{Karimi, D.}, \bibinfo{author}{Samei, G.},
  \bibinfo{author}{Kesch, C.}, \bibinfo{author}{Nir, G.},
  \bibinfo{author}{Salcudean, S.E.}, \bibinfo{year}{2018}.
\newblock \bibinfo{title}{Prostate segmentation in {MRI} using a convolutional
  neural network architecture and training strategy based on statistical shape
  models}.
\newblock \bibinfo{journal}{International journal of computer assisted
  radiology and surgery} \bibinfo{volume}{13}, \bibinfo{pages}{1211--1219}.
%Type = Article
\bibitem[{Kramer et~al.(2020)Kramer, Barkhausen, Bucciarelli-Ducci, Flamm, Kim
  and Nagel}]{journal/JCMR/kramer2020}
\bibinfo{author}{Kramer, C.M.}, \bibinfo{author}{Barkhausen, J.},
  \bibinfo{author}{Bucciarelli-Ducci, C.}, \bibinfo{author}{Flamm, S.D.},
  \bibinfo{author}{Kim, R.J.}, \bibinfo{author}{Nagel, E.},
  \bibinfo{year}{2020}.
\newblock \bibinfo{title}{Standardized cardiovascular magnetic resonance
  imaging ({CMR}) protocols: 2020 update}.
\newblock \bibinfo{journal}{Journal of Cardiovascular Magnetic Resonance}
  \bibinfo{volume}{22}, \bibinfo{pages}{1--18}.
%Type = Article
\bibitem[{Kullback and Leibler(1951)}]{journal/AMS/Kullback1951}
\bibinfo{author}{Kullback, S.}, \bibinfo{author}{Leibler, R.A.},
  \bibinfo{year}{1951}.
\newblock \bibinfo{title}{On information and sufficiency}.
\newblock \bibinfo{journal}{Annals of Mathematical Statistics}
  \bibinfo{volume}{22}, \bibinfo{pages}{79--86}.
%Type = Article
\bibitem[{Lee et~al.(2019)Lee, Petersen, Pawlowski, Glocker and
  Schaap}]{journal/TMI/lee2019}
\bibinfo{author}{Lee, M.C.H.}, \bibinfo{author}{Petersen, K.},
  \bibinfo{author}{Pawlowski, N.}, \bibinfo{author}{Glocker, B.},
  \bibinfo{author}{Schaap, M.}, \bibinfo{year}{2019}.
\newblock \bibinfo{title}{Tetris: Template transformer networks for image
  segmentation with shape priors}.
\newblock \bibinfo{journal}{IEEE transactions on medical imaging}
  \bibinfo{volume}{38}, \bibinfo{pages}{2596--2606}.
%Type = Inproceedings
\bibitem[{Li et~al.(2020a)Li, Weng, Schnabel and Zhuang}]{conf/MICCAI/li2020}
\bibinfo{author}{Li, L.}, \bibinfo{author}{Weng, X.},
  \bibinfo{author}{Schnabel, J.A.}, \bibinfo{author}{Zhuang, X.},
  \bibinfo{year}{2020}a.
\newblock \bibinfo{title}{Joint left atrial segmentation and scar
  quantification based on a {DNN} with spatial encoding and shape attention},
  in: \bibinfo{booktitle}{International Conference on Medical Image Computing
  and Computer-Assisted Intervention}, \bibinfo{organization}{Springer}. pp.
  \bibinfo{pages}{118--127}.
%Type = Article
\bibitem[{Li et~al.(2020b)Li, Wu, Yang, Xu, Wong, Mohiaddin, Firmin, Keegan and
  Zhuang}]{journal/MedIA/li2020}
\bibinfo{author}{Li, L.}, \bibinfo{author}{Wu, F.}, \bibinfo{author}{Yang, G.},
  \bibinfo{author}{Xu, L.}, \bibinfo{author}{Wong, T.},
  \bibinfo{author}{Mohiaddin, R.}, \bibinfo{author}{Firmin, D.},
  \bibinfo{author}{Keegan, J.}, \bibinfo{author}{Zhuang, X.},
  \bibinfo{year}{2020}b.
\newblock \bibinfo{title}{Atrial scar quantification via multi-scale {CNN} in
  the graph-cuts framework}.
\newblock \bibinfo{journal}{Medical Image Analysis} \bibinfo{volume}{60},
  \bibinfo{pages}{101595}.
%Type = Inproceedings
\bibitem[{Li et~al.(2018)Li, Yang, Wu, Wong, Mohiaddin, Firmin, Keegan, Xu and
  Zhuang}]{conf/STACOM/li2018}
\bibinfo{author}{Li, L.}, \bibinfo{author}{Yang, G.}, \bibinfo{author}{Wu, F.},
  \bibinfo{author}{Wong, T.}, \bibinfo{author}{Mohiaddin, R.},
  \bibinfo{author}{Firmin, D.}, \bibinfo{author}{Keegan, J.},
  \bibinfo{author}{Xu, L.}, \bibinfo{author}{Zhuang, X.}, \bibinfo{year}{2018}.
\newblock \bibinfo{title}{Atrial scar segmentation via potential learning in
  the graph-cut framework}, in: \bibinfo{booktitle}{International Workshop on
  Statistical Atlases and Computational Models of the Heart},
  \bibinfo{organization}{Springer}. pp. \bibinfo{pages}{152--160}.
%Type = Article
\bibitem[{Li et~al.(2021a)Li, Zimmer, Schnabel and Zhuang}]{conf/MICCAI/li2021}
\bibinfo{author}{Li, L.}, \bibinfo{author}{Zimmer, V.A.},
  \bibinfo{author}{Schnabel, J.A.}, \bibinfo{author}{Zhuang, X.},
  \bibinfo{year}{2021}a.
\newblock \bibinfo{title}{Atrial{G}eneral: Domain generalization for left
  atrial segmentation of multi-center {LGE MRI}s}.
\newblock \bibinfo{journal}{arXiv preprint arXiv:2106.08727} .
%Type = Article
\bibitem[{Li et~al.(2021b)Li, Zimmer, Schnabel and
  Zhuang}]{journal/MedIA/li2021}
\bibinfo{author}{Li, L.}, \bibinfo{author}{Zimmer, V.A.},
  \bibinfo{author}{Schnabel, J.A.}, \bibinfo{author}{Zhuang, X.},
  \bibinfo{year}{2021}b.
\newblock \bibinfo{title}{Medical image analysis on left atrial {LGE MRI} for
  atrial fibrillation studies: A review}.
\newblock \bibinfo{journal}{arXiv preprint arXiv:2106.09862} .
%Type = Article
\bibitem[{Liu et~al.(2017)Liu, Zhuang, Wu, An, Xu, Peters and
  Gu}]{journal/TBME/liu2017}
\bibinfo{author}{Liu, J.}, \bibinfo{author}{Zhuang, X.}, \bibinfo{author}{Wu,
  L.}, \bibinfo{author}{An, D.}, \bibinfo{author}{Xu, J.},
  \bibinfo{author}{Peters, T.}, \bibinfo{author}{Gu, L.}, \bibinfo{year}{2017}.
\newblock \bibinfo{title}{Myocardium segmentation from {DE MRI} using
  multicomponent {G}aussian mixture model and coupled level set}.
\newblock \bibinfo{journal}{IEEE Transactions on Biomedical Engineering}
  \bibinfo{volume}{64}, \bibinfo{pages}{2650--2661}.
%Type = Article
\bibitem[{Liu et~al.(2019)Liu, Dou, Chen, Qin and Heng}]{journal/tmi/liu2019}
\bibinfo{author}{Liu, L.}, \bibinfo{author}{Dou, Q.}, \bibinfo{author}{Chen,
  H.}, \bibinfo{author}{Qin, J.}, \bibinfo{author}{Heng, P.A.},
  \bibinfo{year}{2019}.
\newblock \bibinfo{title}{Multi-task deep model with margin ranking loss for
  lung nodule analysis}.
\newblock \bibinfo{journal}{IEEE transactions on medical imaging} .
%Type = Article
\bibitem[{Mansoor et~al.(2016)Mansoor, Cerrolaza, Idrees, Biggs, Alsharid,
  Avery and Linguraru}]{journal/TMI/mansoor2016}
\bibinfo{author}{Mansoor, A.}, \bibinfo{author}{Cerrolaza, J.J.},
  \bibinfo{author}{Idrees, R.}, \bibinfo{author}{Biggs, E.},
  \bibinfo{author}{Alsharid, M.A.}, \bibinfo{author}{Avery, R.A.},
  \bibinfo{author}{Linguraru, M.G.}, \bibinfo{year}{2016}.
\newblock \bibinfo{title}{Deep learning guided partitioned shape model for
  anterior visual pathway segmentation}.
\newblock \bibinfo{journal}{IEEE transactions on medical imaging}
  \bibinfo{volume}{35}, \bibinfo{pages}{1856--1865}.
%Type = Article
\bibitem[{Njoku et~al.(2018)Njoku, Kannabhiran, Arora, Reddy, Gopinathannair,
  Lakkireddy and Dominic}]{journal/EP/njoku2018}
\bibinfo{author}{Njoku, A.}, \bibinfo{author}{Kannabhiran, M.},
  \bibinfo{author}{Arora, R.}, \bibinfo{author}{Reddy, P.},
  \bibinfo{author}{Gopinathannair, R.}, \bibinfo{author}{Lakkireddy, D.},
  \bibinfo{author}{Dominic, P.}, \bibinfo{year}{2018}.
\newblock \bibinfo{title}{Left atrial volume predicts atrial fibrillation
  recurrence after radiofrequency ablation: a meta-analysis}.
\newblock \bibinfo{journal}{Ep Europace} \bibinfo{volume}{20},
  \bibinfo{pages}{33--42}.
%Type = Article
\bibitem[{Oktay et~al.(2017)Oktay, Ferrante, Kamnitsas, Heinrich, Bai,
  Caballero, Cook, De~Marvao, Dawes, O‘Regan et~al.}]{journal/TMI/oktay2017}
\bibinfo{author}{Oktay, O.}, \bibinfo{author}{Ferrante, E.},
  \bibinfo{author}{Kamnitsas, K.}, \bibinfo{author}{Heinrich, M.},
  \bibinfo{author}{Bai, W.}, \bibinfo{author}{Caballero, J.},
  \bibinfo{author}{Cook, S.A.}, \bibinfo{author}{De~Marvao, A.},
  \bibinfo{author}{Dawes, T.}, \bibinfo{author}{O‘Regan, D.P.}, et~al.,
  \bibinfo{year}{2017}.
\newblock \bibinfo{title}{Anatomically constrained neural networks ({ACNN}s):
  application to cardiac image enhancement and segmentation}.
\newblock \bibinfo{journal}{IEEE transactions on medical imaging}
  \bibinfo{volume}{37}, \bibinfo{pages}{384--395}.
%Type = Inproceedings
\bibitem[{Perry et~al.(2012)Perry, Morris, Burgon, McGann, MacLeod and
  Cates}]{conf/MI/perry2012}
\bibinfo{author}{Perry, D.}, \bibinfo{author}{Morris, A.},
  \bibinfo{author}{Burgon, N.}, \bibinfo{author}{McGann, C.},
  \bibinfo{author}{MacLeod, R.}, \bibinfo{author}{Cates, J.},
  \bibinfo{year}{2012}.
\newblock \bibinfo{title}{Automatic classification of scar tissue in late
  gadolinium enhancement cardiac {MRI} for the assessment of left-atrial wall
  injury after radiofrequency ablation}, in: \bibinfo{booktitle}{Medical
  Imaging 2012: Computer-Aided Diagnosis}, \bibinfo{organization}{International
  Society for Optics and Photonics}. p. \bibinfo{pages}{83151D}.
%Type = Inproceedings
\bibitem[{Qiao et~al.(2018)Qiao, Wang, van~der Geest and
  Tao}]{conf/STACOM/qiao2018}
\bibinfo{author}{Qiao, M.}, \bibinfo{author}{Wang, Y.},
  \bibinfo{author}{van~der Geest, R.J.}, \bibinfo{author}{Tao, Q.},
  \bibinfo{year}{2018}.
\newblock \bibinfo{title}{Fully automated left atrium cavity segmentation from
  3{D} {GE-MRI} by multi-atlas selection and registration}, in:
  \bibinfo{booktitle}{International Workshop on Statistical Atlases and
  Computational Models of the Heart}, \bibinfo{organization}{Springer}. pp.
  \bibinfo{pages}{230--236}.
%Type = Article
\bibitem[{Ravanelli et~al.(2014)Ravanelli, dal Piaz, Centonze, Casagranda,
  Marini, Del~Greco, Karim, Rhode and Valentini}]{journal/tmi/Ravanelli2014}
\bibinfo{author}{Ravanelli, D.}, \bibinfo{author}{dal Piaz, E.C.},
  \bibinfo{author}{Centonze, M.}, \bibinfo{author}{Casagranda, G.},
  \bibinfo{author}{Marini, M.}, \bibinfo{author}{Del~Greco, M.},
  \bibinfo{author}{Karim, R.}, \bibinfo{author}{Rhode, K.},
  \bibinfo{author}{Valentini, A.}, \bibinfo{year}{2014}.
\newblock \bibinfo{title}{A novel skeleton based quantification and {3-D}
  volumetric visualization of left atrium fibrosis using late gadolinium
  enhancement magnetic resonance imaging}.
\newblock \bibinfo{journal}{IEEE transactions on medical imaging}
  \bibinfo{volume}{33}, \bibinfo{pages}{566--576}.
%Type = Misc
\bibitem[{Rhode and Karim(2012)}]{link/LAScarSeg2012}
\bibinfo{author}{Rhode, K.}, \bibinfo{author}{Karim, R.}, \bibinfo{year}{2012}.
\newblock \bibinfo{title}{{ISBI} 2012: Left atrium fibrosis and scar
  segmentation challenge}.
\newblock
  \bibinfo{howpublished}{\url{http://www.cardiacatlas.org/challenges/left-atrium-fibrosis-and-scar-segmentation-challenge/}}.
%Type = Inproceedings
\bibitem[{Ronneberger et~al.(2015)Ronneberger, Fischer and
  Brox}]{conf/MICCAI/ronneberger2015}
\bibinfo{author}{Ronneberger, O.}, \bibinfo{author}{Fischer, P.},
  \bibinfo{author}{Brox, T.}, \bibinfo{year}{2015}.
\newblock \bibinfo{title}{U-net: Convolutional networks for biomedical image
  segmentation}, in: \bibinfo{booktitle}{International Conference on Medical
  image computing and computer-assisted intervention},
  \bibinfo{organization}{Springer}. pp. \bibinfo{pages}{234--241}.
%Type = Article
\bibitem[{Sugiyama et~al.(2013a)Sugiyama, Kanamori, Suzuki, Plessis, Song and
  Takeuchi}]{journal/NC/Sugiyama2013}
\bibinfo{author}{Sugiyama, M.}, \bibinfo{author}{Kanamori, T.},
  \bibinfo{author}{Suzuki, T.}, \bibinfo{author}{Plessis, M.},
  \bibinfo{author}{Song, L.}, \bibinfo{author}{Takeuchi, I.},
  \bibinfo{year}{2013}a.
\newblock \bibinfo{title}{Density-difference estimation}.
\newblock \bibinfo{journal}{Neural Computation} \bibinfo{volume}{25},
  \bibinfo{pages}{2734--2775}.
%Type = Article
\bibitem[{Sugiyama et~al.(2013b)Sugiyama, Song, Plessis, Yamanaka and
  Kanamori}]{journal/JCSE/Sugiyama2013}
\bibinfo{author}{Sugiyama, M.}, \bibinfo{author}{Song, L.},
  \bibinfo{author}{Plessis, M.}, \bibinfo{author}{Yamanaka, M.},
  \bibinfo{author}{Kanamori, T.}, \bibinfo{year}{2013}b.
\newblock \bibinfo{title}{Direct divergence approximation between probability
  distributions and its applications in machine learning}.
\newblock \bibinfo{journal}{Journal of Computing Science \& Engineering}
  \bibinfo{volume}{7}, \bibinfo{pages}{99--111}.
%Type = Incollection
\bibitem[{Tang et~al.(2017)Tang, Valipour, Zhang, Cobzas and
  Jagersand}]{conf/tang2017}
\bibinfo{author}{Tang, M.}, \bibinfo{author}{Valipour, S.},
  \bibinfo{author}{Zhang, Z.}, \bibinfo{author}{Cobzas, D.},
  \bibinfo{author}{Jagersand, M.}, \bibinfo{year}{2017}.
\newblock \bibinfo{title}{A deep level set method for image segmentation}, in:
  \bibinfo{booktitle}{Deep Learning in Medical Image Analysis and Multimodal
  Learning for Clinical Decision Support}. \bibinfo{publisher}{Springer}, pp.
  \bibinfo{pages}{126--134}.
%Type = Article
\bibitem[{Tao et~al.(2016)Tao, Ipek, Shahzad, Berendsen, Nazarian and van~der
  Geest}]{journal/jmri/Tao2016}
\bibinfo{author}{Tao, Q.}, \bibinfo{author}{Ipek, E.G.},
  \bibinfo{author}{Shahzad, R.}, \bibinfo{author}{Berendsen, F.F.},
  \bibinfo{author}{Nazarian, S.}, \bibinfo{author}{van~der Geest, R.J.},
  \bibinfo{year}{2016}.
\newblock \bibinfo{title}{Fully automatic segmentation of left atrium and
  pulmonary veins in late gadolinium-enhanced {MRI}: Towards objective atrial
  scar assessment}.
\newblock \bibinfo{journal}{Journal of magnetic resonance imaging}
  \bibinfo{volume}{44}, \bibinfo{pages}{346--354}.
%Type = Article
\bibitem[{Tobon-Gomez et~al.(2015)Tobon-Gomez, Geers, Peters, Weese, Pinto,
  Karim, Ammar, Daoudi, Margeta, Sandoval et~al.}]{journal/tmi/tobon2015}
\bibinfo{author}{Tobon-Gomez, C.}, \bibinfo{author}{Geers, A.J.},
  \bibinfo{author}{Peters, J.}, \bibinfo{author}{Weese, J.},
  \bibinfo{author}{Pinto, K.}, \bibinfo{author}{Karim, R.},
  \bibinfo{author}{Ammar, M.}, \bibinfo{author}{Daoudi, A.},
  \bibinfo{author}{Margeta, J.}, \bibinfo{author}{Sandoval, Z.}, et~al.,
  \bibinfo{year}{2015}.
\newblock \bibinfo{title}{Benchmark for algorithms segmenting the left atrium
  from 3{D} {CT} and {MRI} datasets}.
\newblock \bibinfo{journal}{IEEE transactions on medical imaging}
  \bibinfo{volume}{34}, \bibinfo{pages}{1460--1473}.
%Type = Article
\bibitem[{Veni et~al.(2017)Veni, Elhabian and Whitaker}]{journal/mia/Veni2017}
\bibinfo{author}{Veni, G.}, \bibinfo{author}{Elhabian, S.Y.},
  \bibinfo{author}{Whitaker, R.T.}, \bibinfo{year}{2017}.
\newblock \bibinfo{title}{Shapecut: Bayesian surface estimation using
  shape-driven graph}.
\newblock \bibinfo{journal}{Medical image analysis} \bibinfo{volume}{40},
  \bibinfo{pages}{11--29}.
%Type = Inproceedings
\bibitem[{Wu et~al.(2018)Wu, Li, Yang, Wong, Mohiaddin, Firmin, Keegan, Xu and
  Zhuang}]{conf/miccai/wu2018}
\bibinfo{author}{Wu, F.}, \bibinfo{author}{Li, L.}, \bibinfo{author}{Yang, G.},
  \bibinfo{author}{Wong, T.}, \bibinfo{author}{Mohiaddin, R.},
  \bibinfo{author}{Firmin, D.}, \bibinfo{author}{Keegan, J.},
  \bibinfo{author}{Xu, L.}, \bibinfo{author}{Zhuang, X.}, \bibinfo{year}{2018}.
\newblock \bibinfo{title}{Atrial fibrosis quantification based on maximum
  likelihood estimator of multivariate images}, in:
  \bibinfo{booktitle}{International Conference on Medical Image Computing and
  Computer-Assisted Intervention}, \bibinfo{organization}{Springer}. pp.
  \bibinfo{pages}{604--612}.
%Type = Article
\bibitem[{Xiong et~al.(2018)Xiong, Fedorov, Fu, Cheng, Macleod and
  Zhao}]{journal/TMI/xiong2018}
\bibinfo{author}{Xiong, Z.}, \bibinfo{author}{Fedorov, V.V.},
  \bibinfo{author}{Fu, X.}, \bibinfo{author}{Cheng, E.},
  \bibinfo{author}{Macleod, R.}, \bibinfo{author}{Zhao, J.},
  \bibinfo{year}{2018}.
\newblock \bibinfo{title}{Fully automatic left atrium segmentation from late
  gadolinium enhanced magnetic resonance imaging using a dual fully
  convolutional neural network}.
\newblock \bibinfo{journal}{IEEE transactions on medical imaging}
  \bibinfo{volume}{38}, \bibinfo{pages}{515--524}.
%Type = Article
\bibitem[{Xiong et~al.(2020)Xiong, Xia, Hu, Huang, Vesal, Ravikumar, Maier, Li,
  Tong, Si et~al.}]{journal/xiong2020}
\bibinfo{author}{Xiong, Z.}, \bibinfo{author}{Xia, Q.}, \bibinfo{author}{Hu,
  Z.}, \bibinfo{author}{Huang, N.}, \bibinfo{author}{Vesal, S.},
  \bibinfo{author}{Ravikumar, N.}, \bibinfo{author}{Maier, A.},
  \bibinfo{author}{Li, C.}, \bibinfo{author}{Tong, Q.}, \bibinfo{author}{Si,
  W.}, et~al., \bibinfo{year}{2020}.
\newblock \bibinfo{title}{A global benchmark of algorithms for segmenting late
  gadolinium-enhanced cardiac magnetic resonance imaging}.
\newblock \bibinfo{journal}{arXiv preprint arXiv:2004.12314} .
%Type = Inproceedings
\bibitem[{Xue et~al.(2017)Xue, Lum, Mercado, Landis, Warrington and
  Li}]{conf/MICCAI/xue2017}
\bibinfo{author}{Xue, W.}, \bibinfo{author}{Lum, A.}, \bibinfo{author}{Mercado,
  A.}, \bibinfo{author}{Landis, M.}, \bibinfo{author}{Warrington, J.},
  \bibinfo{author}{Li, S.}, \bibinfo{year}{2017}.
\newblock \bibinfo{title}{Full quantification of left ventricle via deep
  multitask learning network respecting intra-and inter-task relatedness}, in:
  \bibinfo{booktitle}{International Conference on Medical Image Computing and
  Computer-Assisted Intervention}, \bibinfo{organization}{Springer}. pp.
  \bibinfo{pages}{276--284}.
%Type = Article
\bibitem[{Xue et~al.(2019)Xue, Tang, Qiao, Gong, Yin, Qian, Huang, Fan and
  Huang}]{conf/xue2019}
\bibinfo{author}{Xue, Y.}, \bibinfo{author}{Tang, H.}, \bibinfo{author}{Qiao,
  Z.}, \bibinfo{author}{Gong, G.}, \bibinfo{author}{Yin, Y.},
  \bibinfo{author}{Qian, Z.}, \bibinfo{author}{Huang, C.},
  \bibinfo{author}{Fan, W.}, \bibinfo{author}{Huang, X.}, \bibinfo{year}{2019}.
\newblock \bibinfo{title}{Shape-aware organ segmentation by predicting signed
  distance maps}.
\newblock \bibinfo{journal}{arXiv preprint arXiv:1912.03849} .
%Type = Article
\bibitem[{Yang et~al.(2020)Yang, Chen, Gao, Li, Ni, Angelini, Wong, Mohiaddin,
  Nyktari, Wage et~al.}]{journal/FGCS/yang2020}
\bibinfo{author}{Yang, G.}, \bibinfo{author}{Chen, J.}, \bibinfo{author}{Gao,
  Z.}, \bibinfo{author}{Li, S.}, \bibinfo{author}{Ni, H.},
  \bibinfo{author}{Angelini, E.}, \bibinfo{author}{Wong, T.},
  \bibinfo{author}{Mohiaddin, R.}, \bibinfo{author}{Nyktari, E.},
  \bibinfo{author}{Wage, R.}, et~al., \bibinfo{year}{2020}.
\newblock \bibinfo{title}{Simultaneous left atrium anatomy and scar
  segmentations via deep learning in multiview information with attention}.
\newblock \bibinfo{journal}{Future Generation Computer Systems}
  \bibinfo{volume}{107}, \bibinfo{pages}{215--228}.
%Type = Article
\bibitem[{Yang et~al.(2018a)Yang, Zhuang, Khan, Haldar, Nyktari, Li, Wage, Ye,
  Slabaugh, Mohiaddin et~al.}]{journal/MP/yang2018}
\bibinfo{author}{Yang, G.}, \bibinfo{author}{Zhuang, X.},
  \bibinfo{author}{Khan, H.}, \bibinfo{author}{Haldar, S.},
  \bibinfo{author}{Nyktari, E.}, \bibinfo{author}{Li, L.},
  \bibinfo{author}{Wage, R.}, \bibinfo{author}{Ye, X.},
  \bibinfo{author}{Slabaugh, G.}, \bibinfo{author}{Mohiaddin, R.}, et~al.,
  \bibinfo{year}{2018}a.
\newblock \bibinfo{title}{Fully automatic segmentation and objective assessment
  of atrial scars for long-standing persistent atrial fibrillation patients
  using late gadolinium-enhanced {MRI}}.
\newblock \bibinfo{journal}{Medical physics} \bibinfo{volume}{45},
  \bibinfo{pages}{1562--1576}.
%Type = Inproceedings
\bibitem[{Yang et~al.(2018b)Yang, Wang, Wang, Wang, Nezafat, Ni and
  Heng}]{conf/STACOM/yang2018}
\bibinfo{author}{Yang, X.}, \bibinfo{author}{Wang, N.}, \bibinfo{author}{Wang,
  Y.}, \bibinfo{author}{Wang, X.}, \bibinfo{author}{Nezafat, R.},
  \bibinfo{author}{Ni, D.}, \bibinfo{author}{Heng, P.A.},
  \bibinfo{year}{2018}b.
\newblock \bibinfo{title}{Combating uncertainty with novel losses for automatic
  left atrium segmentation}, in: \bibinfo{booktitle}{International Workshop on
  Statistical Atlases and Computational Models of the Heart},
  \bibinfo{organization}{Springer}. pp. \bibinfo{pages}{246--254}.
%Type = Inproceedings
\bibitem[{Yue et~al.(2019)Yue, Luo, Ye, Xu and Zhuang}]{conf/MICCAI/yue2019}
\bibinfo{author}{Yue, Q.}, \bibinfo{author}{Luo, X.}, \bibinfo{author}{Ye, Q.},
  \bibinfo{author}{Xu, L.}, \bibinfo{author}{Zhuang, X.}, \bibinfo{year}{2019}.
\newblock \bibinfo{title}{Cardiac segmentation from lge mri using deep neural
  network incorporating shape and spatial priors}, in:
  \bibinfo{booktitle}{International Conference on Medical Image Computing and
  Computer-Assisted Intervention}, \bibinfo{organization}{Springer}. pp.
  \bibinfo{pages}{559--567}.
%Type = Inproceedings
\bibitem[{Zeng et~al.(2019)Zeng, Karimi, Pang, Mohammed, Schneider, Honarvar
  and Salcudean}]{conf/MICCAI/zeng2019}
\bibinfo{author}{Zeng, Q.}, \bibinfo{author}{Karimi, D.},
  \bibinfo{author}{Pang, E.H.}, \bibinfo{author}{Mohammed, S.},
  \bibinfo{author}{Schneider, C.}, \bibinfo{author}{Honarvar, M.},
  \bibinfo{author}{Salcudean, S.E.}, \bibinfo{year}{2019}.
\newblock \bibinfo{title}{Liver segmentation in magnetic resonance imaging via
  mean shape fitting with fully convolutional neural networks}, in:
  \bibinfo{booktitle}{International Conference on Medical Image Computing and
  Computer-Assisted Intervention}, \bibinfo{organization}{Springer}. pp.
  \bibinfo{pages}{246--254}.
%Type = Article
\bibitem[{Zhang et~al.(2020)Zhang, Liu, Wang, Chen, Yuan, Li, Shen, Tang, Chen,
  Xia et~al.}]{journal/MedAI/zhang2020}
\bibinfo{author}{Zhang, J.}, \bibinfo{author}{Liu, M.}, \bibinfo{author}{Wang,
  L.}, \bibinfo{author}{Chen, S.}, \bibinfo{author}{Yuan, P.},
  \bibinfo{author}{Li, J.}, \bibinfo{author}{Shen, S.G.F.},
  \bibinfo{author}{Tang, Z.}, \bibinfo{author}{Chen, K.C.},
  \bibinfo{author}{Xia, J.J.}, et~al., \bibinfo{year}{2020}.
\newblock \bibinfo{title}{Context-guided fully convolutional networks for joint
  craniomaxillofacial bone segmentation and landmark digitization}.
\newblock \bibinfo{journal}{Medical Image Analysis} \bibinfo{volume}{60},
  \bibinfo{pages}{101621}.
%Type = Misc
\bibitem[{Zhao and Xiong(2018)}]{link/LAseg2018}
\bibinfo{author}{Zhao, J.}, \bibinfo{author}{Xiong, Z.}, \bibinfo{year}{2018}.
\newblock \bibinfo{title}{2018 atrial segmentation challenge}.
\newblock \bibinfo{howpublished}{\url{http://atriaseg2018.cardiacatlas.org/}}.
%Type = Article
\bibitem[{Zhuang(2013)}]{journal/jhe/Zhuang2013}
\bibinfo{author}{Zhuang, X.}, \bibinfo{year}{2013}.
\newblock \bibinfo{title}{Challenges and methodologies of fully automatic whole
  heart segmentation: a review}.
\newblock \bibinfo{journal}{Journal of healthcare engineering}
  \bibinfo{volume}{4}, \bibinfo{pages}{371--407}.
%Type = Article
\bibitem[{Zhuang(2019)}]{journal/PAMI/zhuang2019}
\bibinfo{author}{Zhuang, X.}, \bibinfo{year}{2019}.
\newblock \bibinfo{title}{Multivariate mixture model for myocardial
  segmentation combining multi-source images}.
\newblock \bibinfo{journal}{IEEE transactions on pattern analysis and machine
  intelligence} \bibinfo{volume}{41}, \bibinfo{pages}{2933--2946}.
%Type = Article
\bibitem[{Zhuang et~al.(2019)Zhuang, Li, Payer, {\v{S}}tern, Urschler,
  Heinrich, Oster, Wang, Smedby, Bian et~al.}]{journal/MedAI/zhuang2019}
\bibinfo{author}{Zhuang, X.}, \bibinfo{author}{Li, L.}, \bibinfo{author}{Payer,
  C.}, \bibinfo{author}{{\v{S}}tern, D.}, \bibinfo{author}{Urschler, M.},
  \bibinfo{author}{Heinrich, M.P.}, \bibinfo{author}{Oster, J.},
  \bibinfo{author}{Wang, C.}, \bibinfo{author}{Smedby, {\"O}.},
  \bibinfo{author}{Bian, C.}, et~al., \bibinfo{year}{2019}.
\newblock \bibinfo{title}{Evaluation of algorithms for multi-modality whole
  heart segmentation: An open-access grand challenge}.
\newblock \bibinfo{journal}{Medical image analysis} \bibinfo{volume}{58},
  \bibinfo{pages}{101537}.

\end{thebibliography}

\end{document}